\documentclass[final,titlepage,oneside,12pt]{article}

\def\pm{{{+}\atop{-}}}

\newtheorem{theorem}{Theorem}

\begin{document}

\title{String without strings}
\author{James T. Wheeler \\
{\it {Department of Physics, Utah State University, Logan, UT 84322}}\\
jwheeler@cc.usu.edu}

\maketitle

\begin{abstract}
Scale invariance provides a principled reason for the physical importance
of Hilbert space, the Virasoro algebra, the string mode expansion,
canonical commutators and Schroedinger evolution of states, independent of
the assumptions of string theory and quantum theory. The usual properties
of dimensionful fields imply an infinite, projective tower of conformal
weights associated with the tangent space to scale-invariant spacetimes.
Convergence and measurability on this tangent tower are guaranteed using a
scale-invariant norm, restricted to conformally self-dual vectors. Maps on
the resulting Hilbert space are correspondingly restricted to semi-definite
conformal weight. We find the maximally- and minimally-commuting, complete
Lie algebras of definite-weight operators. The projective symmetry of the
tower gives these algebras central charges, giving the canonical commutator
and quantum Virasoro algebras, respectively. Using a continuous,
m-parameter representation for rank-m tower tensors, we show that the
parallel transport equation for the momentum vector of a particle is the
Schroedinger equation, while the associated definite-weight operators obey
canonical commutation relations. Generalizing to the set of integral curves
of general timelike, self-dual vector-valued weight maps gives a lifting
such that the action of the curves parallel transports arbitrary tower
vectors. We prove that the full set of Schroedinger-lifted integral curves
of a general self-dual map gives an immersion of its 2-dim parameter space
into spacetime, inducing a Lorentzian metric on the parameter space. This
immersion is shown to satisfy the full variational equations of open string.
\end{abstract}


\baselineskip=16pt

\section{Introduction}

\smallskip

It is certainly fair to say that the string mode operators and the
associated Virasoro algebra, acting on Hilbert space, form the heart and
soul of string theory. States can be classified by mass and spin because we
can locally construct generators of the Poincar\'{e} group from the mode
operators; this is also the reason the effective spin-2 field equations must
be generally covariant. The action of the Virasoro algebra determines the
space of physical states, and therefore is responsible for the spin-2
(gravitational) mode present in the closed string ground state. Having a
generally covariant, massless, spin-2 mode is a necessary (though not
sufficient) condition for string theory to approximate general relativity.
Upon quantization, it is the central charge of the Virasoro algebra that
fixes the dimension of the target space at 26 in the bosonic theory or 10 in
the supersymmetric case. And it is the existence of supersymmetric and
heterotic representations of the basic symmetries that frees string theory
from tachyons, guarantees fermions and provides a large internal nonabelian
gauge symmetry.

It is therefore remarkable to discover that exactly this configuration - the
mode algebra and the Virasoro algebra with central charges, acting on
Hilbert space - \textit{occurs classically in the tangent space to every
scale-invariant geometry.} Hilbert space occurs as the tower of conformal
weights, while the mode and Virasoro algebras occur as the unique complete
Lie algebras of maximally or minimally commuting, definite-weight operators.

The missing element in recognizing this fact earlier is the failure to see
the extended character of the tangent space of scale invariant geometries.
The different scaling behavior of fields of different dimension leads to the
association of a \textit{conformal weight} with each field. These conformal
weights are associated with an equivalence relation on the tangent space at
each point of spacetime. The usual Minkowski tangent space modulo this
equivalence relation gives the \textit{tangent tower}, comprised of a copy
of projective Minkowski space for each possible weight. The tangent tower
has a natural scale-invariant inner product under which the conformally
self-dual \textit{weight vectors} form a Hilbert space. A study of linear
operators on the tangent tower reveals the presence of the mode operator and
Virasoro algebras, while the existence of nontrivial projective
representations gives rise to central charges.

While the spontaneous breaking of conformal invariance in field theory is
well known, the scaling symmetry described here remains \textit{unbroken.}
To see clearly the difference between the two symmetries, consider a typical
example of spontaneous symmetry breaking. It is common for an initially
scale-invariant Lagrangian to develop one-loop corrections of the general
form
\begin{equation}
\lambda ^{2}\ln (\frac{\Lambda }{\mu })
\end{equation}
where $\mu $ is fixed by a particle mass. Often, the necessary inclusion of
the dimensionful parameter $\Lambda $ breaks the original scaling symmetry.
However, the dimensionality of $\Lambda $ is determined by a much deeper,
unbreakable symmetry - the matching of the units of $\Lambda $ to the units
of $\mu .$ Reliance on the inviolable balance of units is how we knew the
dimensions of $\Lambda $ to begin with. It is this balance of units which we
refer to here as scale-invariance, a symmetry without which no physical
equation makes sense. Thus, in the ``conformal symmetry breaking''
logarithmic expression above, the transformation $\Lambda \rightarrow
e^{\varphi }\Lambda ,\mu \rightarrow e^{\varphi }\mu $ is a \textit{distinct}
scale symmetry which remains unbroken. It is remarkable that the simple fact
of using units to describe physical quantities implies the rich formalism of
the weight tower.

The layout of our investigation of these ideas is as follows. In Sec.(2), we
present a very brief review of some core elements of string theory. Next, in
Sec.(3), we discuss the properties of scale-invariant geometries, giving the
simple example of a Weyl geometry for a typical context, but focussing
principally on the tangent space structure common to all scaling geometries.
After defining the tangent tower we study its linear transformations in
order to define tensors on the tower. Restricting to
definite-conformal-weight linear transformations as dictated by the
requirements of measurement, we seek a Lie algebra of operators. Two are
found: the maximally commuting and maximally non-commuting algebras
containing one operator of each weight. The commuting algebra is trivial to
derive, but a detailed proof is required to show that the maximally
non-commuting, complete Lie algebra of definite-weight operators is the
Virasoro algebra. The details of the proof are given in Sec.(4).

Once these algebraic structures are specified, we look at a continuous
representation for tower tensors, using $n$ continuous, bounded variables
for rank-$n$ tower tensors. We show that the zero-weight and
zero-anti-weight operators together span the space of harmonic functions and
find continuous representations for the Virasoro and mode algebras. Then, in
Sec.(6), we look first at the effects of different possible equivalence
relations in defining the tower structure, then at the effects of the
projective group representations permitted by the tower symmetry, including
central extensions for both the mode and Virasoro algebras.

As preparation for our study of the conformal dynamics of stringlike
variables, the conformal dynamics of a point particle is sketched in
Sec.(7). We show how parallel transport on the tangent tower gives the
Schrodinger equation for the evolution of tower vectors. We also show how
the self-duality required for convergence of tower vectors combines with the
projective equivalence that defines the tower to give canonical commutation
relations for conformally conjugate variables.

The results of Secs.(5),(6) and (7) are used in Sec.(8) to show an important
relationship of vector-valued weight maps ($(1,2)$ tensor fields) to
strings: the integral transport curves of a timelike, self-dual $(1,2)$
tensor field provide an immersion of $(0,\pi )\times (0,\pi )$ into
spacetime, causal in the pull-back of the spacetime metric, such that the
immersed $2$-surface extremizes the Polykov string action with respect to
variations of the string coordinates, the string endpoints, and the $2$%
-metric.

Sec.(9) concludes the paper with a concise summary.

Throughout our development, we take the point of view that all steps should
be motivated from the principle of scale invariance, even though the
structures which emerge are already familiar from quantum systems. Thus,
scale-invariance provides an underlying reason, within the context of a
spacetime manifold, for the physical existence and importance of Hilbert
space, harmonic operators and the mode and Virasoro algebras which is
independent of the insights of string theory and quantum theory. We use the
name \textit{conformal dynamics} to describe this new point of view.

\section{The String Algebra}

We first provide a brief reminder of some basic features of string and the
Virasoro algebra. Following [1], the construction of the algebra of string
begins with the classical mode expansion. For simplicity we will restrict
our remarks to open string, since the results for closed string are similar
and readily available elsewhere [1].

The classical mode expansion for string, after imposing open string boundary
conditions, is
\begin{equation}
X^{\mu }(\sigma ,\tau )=x^{\mu }+l^{2}p^{\mu }\tau +il\sum_{n>0}\frac{1}{n}\
\alpha _{n}^{\mu }\cos n\sigma \ e^{-in\tau }  \label{Open string}
\end{equation}
where the length parameter $l$ is related to the string tension by $l=(\pi
T)^{-1/2}$. Here $x^{\mu }$ is the center of mass of the string and $p^{\mu
} $ is the total linear momentum. The angular momentum of the string may be
written in terms of $x^{\mu },p^{\mu }$ and the remaining mode amplitudes $%
\alpha _{n}^{\mu }$ as
\begin{equation}
M_{\mu \nu }=x^{\mu }p^{\nu }-x^{\nu }p^{\mu }-i\sum_{n=1}^{\infty }\frac{1}{%
n}(\alpha _{-n}^{\mu }\alpha _{n}^{\nu }-\alpha _{-n}^{\nu }\alpha _{n}^{\mu
})
\end{equation}
Notice that eq.(\ref{Open string}) does \textit{not} have independent left-
and right-moving modes, but only supports the wave modes of the form
\begin{equation}
e^{im(x-y)}+e^{-im(x+y)}
\end{equation}
Similarly, the closed string boundary conditions restrict the solution to
even modes. We shall see that this counting is important for understanding
string from the point of view of scale invariance.

Also from the mode amplitudes, we can construct the quantities,
\begin{equation}
L_{m}=\frac{1}{2}\sum\limits_{-\infty }^{\infty }\alpha _{m-n}\cdot \alpha
_{n}
\end{equation}
which form the Virasoro algebra
\begin{equation}
\lbrack L_{m},L_{n}]_{P.B.}=(m-n)L_{m+n}
\end{equation}
where $[L_{m},L_{n}]_{P.B.}$ denotes the Poisson bracket with respect to the
set of mode coordinates $(x^{\mu },p^{\mu },\alpha _{n}^{\mu })$. To
quantize the
string, a Hilbert space is introduced and the mode amplitudes $\alpha
_{n}^{\mu }$ become Hilbert space operators $\hat{\alpha}_{n}^{\mu },$ $\hat{
\alpha}_{0}^{\mu }=l\hat{p}^{\mu }.$ These operators comprise a countable
set of creation and anihilation operators for modes of vibration of the
string. Also, central charges, $c_{mn}=\frac{D}{12}(m^{3}-m)\delta
_{m+n}^{0} $, may be added to the quantum commutators because the operators $%
\hat{\alpha}_{m}^{\mu }$ are defined only up to a phase, that is, they form
a projective representation of the string mode algebra. The quantum string
algebra is therefore
\begin{equation}
\lbrack \hat{L}_{m},\hat{L}_{n}]=(m-n)\hat{L}_{m+n}+\frac{D}{12}%
(m^{3}-m)\delta _{m+n}^{0}  \label{Vira}
\end{equation}
Quantization also leads to a normal ordering ambiguity in the definition of $%
\hat{L}_{0},$ which is conveniently defined as
\begin{equation}
\hat{L}_{0}\equiv \frac{1}{2}\hat{\alpha}_{0}^{2}+\frac{1}{2}%
\sum\limits_{n=1}^{\infty }\hat{\alpha}_{-n}\hat{\alpha}_{n}
\end{equation}

We can extend the Virasoro algebra further by including symmetries generated
by the stress-energy tensor of the string. The extension takes the form
\begin{eqnarray}
\lbrack \hat{M}^{\alpha \beta },\hat{P}_{n}^{\gamma }] &=&\eta ^{\beta
\gamma }\hat{P}_{n}^{\alpha }-\eta ^{\alpha \gamma }\hat{P}_{n}^{\beta } \\
\lbrack \hat{P}_{m}^{\alpha },\hat{P}_{n}^{\nu }] &=&0 \\
\lbrack \hat{P}_{m}^{\alpha },\hat{L}_{n}] &=&(m-n)\hat{P}_{m+n}^{\alpha }
\label{Momentum extension}
\end{eqnarray}
A similar extension of the Virasoro symmetry follows from the symmetry of
the tangent tower, by considering inhomogeneous linear maps [2], though we
will not go into the details of this extension here.

Finally, the physical space of states $\left| \psi \right\rangle $ is given
by the conditions:
\begin{eqnarray}
\hat{L}_{m}\left| \psi \right\rangle &=&0\qquad \forall m>0 \\
(\hat{L}_{0}-a)\left| \psi \right\rangle &=&0
\end{eqnarray}
These conditions guarantee that physical states have positive norm, and
therefore form a Hilbert space.

\section{The symmetry of scale invariant theories}

We now begin our demonstration that Hilbert space, acted on by mode
amplitudes and the Virasoro algebra \textit{including the central charge},
occurs \textit{classically }as part of the tangent space structure to any
scale-invariant theory of gravity. The mode and Virasoro algebras occur as
the maximally and minimally commuting, complete Lie algebras of linear
transformations of the Hilbert space sector of the tangent space.

To provide the concreteness of a typical background context, we give the
structure equations for the simplest class of scale-invariant geometries.
The only property of these geometries of interest to us here is the tower
structure of the tangent space. Many other geometries besides the class
presented here (for example, conformal or biconformal geometries) have the
same tangent space tower structure.

As the example context, consider a $d$-dim pseudo-Riemannian geometry with
curvature 2-form $\mathbf{R}_{b}^{a}$, vielbein 1-form $\mathbf{e}^{a},$ and
spin connection 1-form $\mathbf{\omega }_{b}^{a}$,  satisfying
\begin{eqnarray}
\mathbf{de}^{a} &=&\mathbf{e}^{b}\wedge \mathbf{\omega }_{b}^{a}
\label{structure equations} \\
\mathbf{R}_{b}^{a} &=&\mathbf{d\omega }_{b}^{a}+\mathbf{\omega }%
_{d}^{a}\wedge \mathbf{\omega }_{b}^{d}  \label{curvature}
\end{eqnarray}
where $a,b=0,1,\ldots d-1.$ The vielbein provides an orthonormal basis, so
that Latin indices are contracted using $\eta _{ab}.$ Conversion between
orthonormal and general coordinate bases is accomplished using the vielbein
component matrix, $e_{\mu }^{\quad a}$ and its inverse.

We can make this geometry scale-invariant by introducing a gauge 1-form $%
\mathbf{\theta }$ (the ``Weyl vector'') for local scale changes together
with its curvature 2-form
\begin{equation}
\Omega =\mathbf{d\theta }  \label{Dilational curvature}
\end{equation}
Unlike Weyl's theory, this curvature is related to the Hamiltonian structure
of fields on spacetime [3]. In particular, $\Omega $ does \textit{not}
represent the electromagnetic field.

The metric is given in terms of the components of the vielbein by
\begin{equation}
g_{\alpha \beta }=e_{\alpha }^{\quad a}e_{\beta }^{\quad b}\eta _{ab}
\end{equation}
where $\eta _{ab}$ is the $d$-dim Minkowski metric. When a local change in
the choice of units changes $g_{\alpha \beta }$ by a (conventional) factor $%
\exp (2\varphi )$, we (conventionally) take the vielbein to scale by $\exp
(\varphi )$%
\begin{equation}
\mathbf{e}^{a}\longrightarrow \exp (\varphi )\ \mathbf{e}^{a}
\end{equation}
The metric is said to have a scale weight of two, and the vielbein a scale
weight of one. In addition, we must modify eq.(\ref{structure equations}) to
give the dependence of the connection on $\mathbf{\theta }$. We have
\begin{equation}
\mathbf{de}^{a}=\mathbf{e}^{b}\wedge \mathbf{\omega }_{b}^{a}+\mathbf{\,e}%
^{a}\wedge \mathbf{\theta }  \label{e sturcture eq in Weyl}
\end{equation}
Eqs.(\ref{curvature},\ref{Dilational curvature} \& \ref{e sturcture eq in
Weyl}) describe scale invariant (or Weyl) geometry. We will not develop
these equations further here, but instead turn our attention to the tangent
space $V$ at a point $P$ of the Weyl, or other scale-invariant, geometry.

\textit{A priori,} the vectors in $V$ transform under the Weyl group, i.e.,
Lorentz transformations and scalings. Translations may also be allowed,
depending on the model. However, the situation is more complicated than
this. The generator of scale changes, $D,$ called the dilation operator, may
have different eigenvalues when acting on different physical fields. These
eigenvalues, called the \textit{conformal weights} of the fields, give rise
to equivalence classes of vectors. In Sec.(6) we show that there are
actually two different equivalence relations consistent with the weight
tower structure, with distinct physical consequences. In each case, the
total tangent vector space at a point $P$ is the direct sum of these
equivalence classes.

Thus, $V$ is a space of vectors $v_{(w)}^{\quad a},$ where in addition the
Lorentz index $a$, each vector carries a weight $w\in W$ which characterizes
its transformation under scalings. Closure of the symmetry algebra under
commutation requires the weight set $W$ to be closed under the addition of
any two \textit{different }elements of $W$. In addition to the integers, $J$%
, the rationals, $Q$, and the reals, $R,$ this constraint permits the finite
sets $\{0,1\}$ and $\left\{ -1,0,1\right\} $. For the remainder of our
discussion, we will assume\footnote{%
Almost all physical systems allow the choice $W=J.$ To arrive at a set of
integer weights $W$ for physical fields, we may use the following procedure.
First, using the fundamental constants, express the units of each given
physical field as a power of length. The set $S$ of all such powers for a
given field theory is typically a set of rational numbers with a least
common denominator, $r$. The set of realized physical weights is then $%
rS\subseteq J$ and we can take the integers $J$ as the weight set. In this
case, the weight of any given physical field is $r$ times the power of
length associated with that field. Only if $S$ has no least common
denominator must the set of weights be extended to (at least) the set of
rationals, $Q.$} that the weight set is $J.$

When the vielbein is scaled by $\exp \varphi $, a vector field of weight $n$
scales by $\exp n\varphi $:
\begin{equation}
v_{(n)}^{a}\longrightarrow e^{n\varphi }\ v_{(n)}^{a}
\end{equation}
Infinitesimally, we have
\begin{equation}
Dv_{(n)}^{a}=nv_{(n)}^{a}
\end{equation}
where $D$ is the generator of dilations. Thus, the covariant derivative of a
weight-$n$ vector field is
\begin{equation}
D_{\mu }v_{(n)}^{a}=\partial _{\mu }v_{(n)}^{a}-v_{(n)}^{b}\omega _{\mu
b}^{\quad a}+v_{(n)}^{a}n\theta _{\mu }  \label{Weyl cov. derivative}
\end{equation}
where $\theta _{\mu }$ is the Weyl vector. Notice how there is effectively a
different gauge vector, $n\theta _{\mu }$, for each weight of vector field.
We handle this multiplicity more explicitly and efficiently by introducting
the tangent tower. The \textit{tangent tower} is the direct sum of vector
spaces $V_{(n)}$ of each allowed weight $n\in J$
\begin{equation}
V=\bigoplus_{n\in J}V_{(n)}
\end{equation}

Now consider homogeneous\footnote{%
As mentioned in Sec.(2), the \textit{inhomogeneous} maps on $V$ lead to the
momentum extension of the Virasoro algebra given by eqs.(7-9).} linear maps
on $V$. These maps must preserve the direct sum structure of $V$, and
therefore must map vectors of definite weight to other vectors of definite
weight.
\begin{equation}
v_{(n)}^{a\ \prime }=\Delta _{(n,m)\quad b}^{\quad \quad a}\,v_{(m)}^{b}
\end{equation}
Here the labels $(n,m)$ on $\Delta _{(mn)\quad b}^{\quad \quad a}$ are just
a mnemonic to tell us that $\Delta _{(mn)\quad b}^{\quad \quad a}$ carries
the weight-$m$ vector to a weight-$n$ vector.

The form of $\Delta _{(mn)\quad b}^{\quad \quad a}$ is restricted by
requiring consistency with the usual procedures for combining physical
fields, which let us construct vector fields of all weights from a single
vector field of weight $n$. This is simple, for given a nonvanishing vector
field\footnote{%
Any time orientable spacetime manifold has a nonvanishing vector field.} $%
v_{(n)}^{a}$ of weight $n,$ and the metric of weight $2,$ the nonvanishing
unit weight scalar field
\begin{equation}
\phi _{(1)}\equiv (\eta _{ab}v_{(n)}^{a}\,v_{(n)}^{b})^{1/(2n+2)}
\label{weighted scalar}
\end{equation}
can be used multiplicatively to produce vector fields of arbitrary weight,
\begin{equation}
v_{(n+k)}^{a}=(\phi _{(1)})^{k}v_{(n)}^{a}.
\end{equation}
Clearly, $v_{(n+k)}^{a}$ and $v_{(n)}^{a}$ are parallel. Moreover, given
this construction, it is clear that at any point the angle between $%
v_{(n)}^{a}$ and a second similar set of vector fields $w_{(n)}^{a}$, is the
same as that between $v_{(n+k)}^{a}$ and $w_{(n+k)}^{a}$ independent of the
weight $n$ and the value of $k$. Therefore, the Lorentz transformations
contained in $\Delta _{(mn)\quad b}^{\quad \quad a}$ must be independent of
weight, and $\Delta _{(mn)\quad b}^{\quad \quad a}$ is a direct product
\begin{equation}
\Delta _{(mn)\quad b}^{\quad \quad a}=\Delta _{(nm)}\Lambda _{\quad b}^{a}
\end{equation}
where $\Lambda _{\quad b}^{a}$ is a Lorentz transformation and $\Delta %
_{(nm)}$ acts only on the weight indices.

Now consider $\mathcal{T}:V^{n}\rightarrow R,$ the space of multilinear maps
over $V=\oplus V^{n}$, typified by sums over a multi-vector basis,
\begin{equation}
T_{n_{1}+n_{2}+\cdots n_{3}}^{ab...c}=\sum u_{(n_{1})}^{\quad
a}v_{(n_{2})}^{\quad b}\cdots w_{(n_{3})}^{\quad c}
\label{vector based tensors}
\end{equation}
Though $\mathcal{T}$ seems to be the space of rank-$n$ tensors over $V,$ eq.(%
\ref{vector based tensors}) is actually too restricted. Because of the
separation of Lorentz transformations and weight transformations, we can
factor apart the direction and weight properties of the vectors, and use the
factors to construct tensors of arbitrarily mixed type. The Lorentz
direction factor will be an element of $d$-dim, projective Minkowski space, $%
PM^{d}\equiv M^{d}/R^{+}.$ The weight factor, $\Delta _{(mn)},$ acts on the
integers, $J,$ so we consider tensors of type $\left( r,s\right) $ to be
linear maps $T^{(r,s)}:$ $(PM^{d})^{r}\otimes J^{s}\rightarrow R.$ Such a
tensor, of rank $\left( r,s\right) ,$ will have components of the form
\begin{equation}
T_{n_{1}n_{2}\cdots n_{s}}^{a_{1}a_{2}\cdots a_{r}}
\end{equation}
where in contrast to eq.(\ref{vector based tensors}), we may have $r\neq s.$
For example, the $(0,s)$ tensor $T_{n_{1}\cdots n_{s}}$ is a multilinear map
on an $s$-tuple of rank $(0,1)$ fields, $(\varphi _{(k_{1})},\ldots ,\varphi
_{(k_{s})}).$ Rank $(0,1)$ fields are Lorentz scalars and weight tower
vectors. Tower tensors of rank $(r,s)$ must not be confused with tensors of
mixed covariant and contravariant types. Once the metric is introduced, both
the Lorentz rank $r,$ and the weight rank $s,$ may be subdivided among
covariant or contravariant types.

It is important to distinguish the conformal weight of a tensor from its
rank over weight space. Each $(r,s)$ tensor $T^{(r,s)}$ acts on $s$ fields
of definite or indefinite weight. If we evaluate $T^{(r,s)}$ on a set of
fields $\phi _{1},\cdots ,\phi _{s}$ of definite weights $n_{1},\cdots
,n_{s} $ respectively, then $T^{(r,s)}$ is of weight $k$ if the weight of $%
T^{(r,s)}(\phi _{1},\cdots ,\phi _{s})$ is $k+n_{1}+\cdots +n_{s},$
independent of the values of $n_{1},\cdots ,n_{s}.$ Such a tensor $T^{(r,s)}$
has tower rank $s$ and conformal weight $k.$ It is possible for $T^{(r,s)}$
to have no weight, or to have a weight only with respect to certain of its
indices.

Our principal concerns here will be with Lorentz tensors of weight ranks $%
s=0,1$ and $2.$ In many instances the Lorentz rank is immaterial, though it
is often simplest to consider Lorentz scalars or vectors, $r=0,1.$ We now
consider each of these values of $s$ in turn.

Tensors of type $\left( r,0\right) $ are rank-$r$ Lorentz tensors, $%
T^{a_{1}a_{2}\cdots a_{r}}=T_{0}^{a_{1}a_{2}\cdots a_{r}}$ which are
annihilated by the dilation operator
\begin{equation}
DT^{a_{1}a_{2}\cdots a_{r}}=0
\end{equation}
Of these, the Lorentz scalars are centrally important. This is because
\textit{type }$(0,0)$\textit{\ tensors are the only physically measurable
quantities}, all others changing under conformal transformations, Lorentz
transformations, or both. Tensors with $s\neq 0$ are useful, however, for
the construction of weight zero tensors, just as Lorentz tensors are useful
for constructing Lorentz scalars.

Type $(0,1)$ tensors are weighted Lorentz scalars such as $\phi _{(1)}$
defined in eq.(\ref{weighted scalar}), or their sums. To express higher
weight objects we can use powers of $\phi _{(1)},$ such as
\begin{equation}
\phi _{(k)}\equiv \left( \phi _{(1)}\right) ^{k}
\end{equation}
as an example of a scalar with weight $k,$ i.e., $D\phi _{(k)}=k\phi _{(k)}$
. Taking a fixed set $\{\phi _{(k)}\}$ as a \textit{vector} basis for the
type $(0,1)$ tensors, we see that the general element
\begin{equation}
f(\phi )=\sum\limits_{k=0}^{\infty }\beta _{k}\phi _{(k)}
\end{equation}
may be written as an arbitrary convergent power series in $\phi ,$ of
indefinite weight. Linear combinations of fields of different scale weights
are unphysical in themselves, but may nonetheless provide measurable
quantities. For example, while $f(\phi )$ is not measurable directly, the $0$%
-weight scalar
\begin{equation}
\sum\limits_{k=0}^{\infty }\beta _{k}\beta _{-k}
\end{equation}
constructible from $f(\phi )$ can in principle be measured. In Sec.(5), the
allowed class of such indefinite weight superpositions will be defined.

The tensors of type $(1,1)$ are simply the vectors $v_{n}^{a}$ we began
with, together with their linear combinations. Again, the idea of taking
linear combinations of lengths and areas seems physically odd, but as long
as we remember that only type $(0,0)$ objects are measurable, there is no
inherent problem.

Finally, we will be particularly interested in objects of type $(r,2)$
because they act as linear maps on the weight labels of type $(r^{\prime },1)
$ objects. Our first result below gives a set of necessary and sufficient
conditions for the class of type $(0,2)$ objects, called \textit{weight maps,%
} to reduce to the Virasoro algebra, while Theorem 2 similarly characterizes
the mode algebra. The type $(1,2)$ objects are vector-valued weight maps
including a subset analogous to the quantized Fourier coefficients $\hat{%
\alpha}_{n}^{a}$ of strings. We explore the properties of these in detail in
Sec.(8).

\smallskip

We now begin our study of weight maps. Consider possible maps on the set of
integer labels $\mathcal{M}:J\longrightarrow J,$ that is, type $(0,2)$
tensors. Together with the generators of the Weyl group, we demand that
these operators have the following properties:

\begin{enumerate}
\item  The operators form a Lie algebra.

\item  All operators have definite weight.

\item  The algebra is maximally non-commuting, i.e., $[M_{(m),}M_{(n)}]$ $%
\neq 0$ for all $m\neq n.$

\item  The algebra is complete, containing exactly one operator of each
allowed weight.
\end{enumerate}

Assumptions (1) and (2) are basic physical constraints. The demand for an
algebra instead of a group follows from the basic commutator relation $%
[D,M_{(n)}]$ $=nM_{(n)},$ which shows that definite weight operators extend
the Lorentz-plus-dilation Lie algebra. Assumption (3) avoids trivial
subgroups of operators and isomorphic representations.\footnote{%
For example, the set of even integers is an allowed weight set isomorphic to
$J.$} Though it is clearly possible to find algebras with multiple operators
or no operators of a given weight, condition (4) is a reasonable
completeness demand.

The remainder of Sec.(3) and all of Sec.(4) together provide the proof of
the following central result. Sec.(3) includes the more general elements of
the proof, as well as two additional theorems, while Sec.(4) is devoted to
the detailed inductive analysis required by the proof of Theorem 1.

\smallskip

\begin{theorem}
The unique maximally non-commuting, complete Lie algebra of definite-weight
operators is the Virasoro algebra.
\end{theorem}

To make the flow of the argument clearer, we choose a particular matrix
basis for the next steps. While the matrix basis we use in this section
provides an intuitive handle on the algebra, other representations
generalize readily to other rank objects, allow easier calculation and
reveal new properties of the objects. Therefore, in Sec.(5) we will
introduce an alternative representation.

To construct the matrix basis, we pick an arbitrary definite-weight vector $%
v_{(n)}^{a}$ in $V$. By the construction above, we can produce an infinite
tower of parallel vectors of each weight $m\in J$ . Because Lorentz
transformations decouple from weight transforms, we need only consider maps
which take some vector $v_{(n)}^{a}$ in this tower to some other $v_{(m)}^{a}
$ in the tower. Therefore, we can suppress the Lorentz index and choose a
basis in which the components of $v_{(n)}^{a}=v_{(n)}$ are given by
\begin{equation}
(v_{(n)}^{a})^{i}=v^{a}\delta _{n}^{i}
\end{equation}
or simply
\begin{equation}
(v_{(n)})^{i}=\delta _{n}^{i}
\end{equation}
Alternatively, we can work with $(0,1)$ tensors, with a definite-weight
basis $\phi _{(n)},$ and avoid Lorentz structure altogether. In this case,
the basis is essentially the same, ($\phi _{(n)})^{i}=\delta _{n}^{i},$ but
there is no suppressed index\footnote{%
Ultimately, it makes sense to identify all of the ranks $(1,0)$, $(1,1)$ and
$(0,1)$ as vectors in $V$, with $(1,0)$ vectors being those that lie on a
single ``floor'' of the tower and $(0,1)$ vectors having the zero Lorentz
vector as their projection to any floor. Although we do not introduce a
combined basis notation covering all types, we will refer to all three types
as vectors.}.

Now consider those linear maps that associate to each definite-weight vector
another definite-weight vector. We can describe this class as follows. Let
\begin{equation}
\Phi =\{\phi \mid \phi :J\rightarrow J\}
\end{equation}
be the automorphism group of the integers. Then the weight tower is
preserved by the set of maps
\begin{equation}
\mathcal{M}^{\Phi }=\{M^{\phi ,\alpha }\mid M^{\phi ,\alpha }v_{(k)}=\alpha
_{k}^{\phi }v_{\phi (k)},\phi \in \Phi \}
\end{equation}
where for each $M^{\phi ,\alpha }$ the set $\{\alpha _{k}^{\phi }\mid k\in
J,\alpha _{k}^{\phi }\in R\}$ may be chosen arbitrarily.

The action of $D$ on the set $\mathcal{M}^{\Phi }$ is not well-defined
because there is no weight associated with a general map $M^{\phi }.\ $For
this reason, and because well-defined weight is necessary for constructing
conformal scalars, we restrict our attention to maps of definite weight.
Definite-weight maps are defined by using the additivity property of
conformal weights. Consider the action of $D$ on a mapping $M^{\phi }$ $%
v_{(k)}$ of a weight-$k$ vector
\begin{equation}
DM^{\phi }v_{(k)}=[D,M^{\phi }]v_{(k)}+M^{\phi }Dv_{(k)}
\end{equation}
Solving for the commutator and using $M^{\phi }v_{(k)}=\alpha _{k}v_{(\phi
(k))}$ results in
\begin{equation}
\lbrack D,M^{\phi }]v_{(k)}=(\phi (k)-k)\alpha _{k}v_{(\phi (k))}=(\phi
(k)-k)M^{\phi }v_{(k)}
\end{equation}
The definite-weight weight maps are those for which this is an operator
relation,
\begin{equation}
\lbrack D,M^{\phi }]=(\phi (k)-k)M^{\phi }  \label{DM commutator}
\end{equation}
which requires that the coefficient $(\phi (k)-k)$ be independent of $k.$
Since $\phi (k)\in J,$ this means there exists some integer $n$ such that $%
\phi (k)=n+k$ so the maps $\phi $ are restricted to the countable set
\begin{equation}
\Phi _{N}=\{\phi _{n}\mid \phi _{n}(k)=n+k\}\subset \Phi
\end{equation}
where $n\equiv \phi (0).$ Therefore, we can denote the set of
definite-weight operators by
\begin{equation}
\mathcal{M}^{N,A}=\{M^{n,\alpha }\mid M^{n,\alpha }v_{(k)}=\alpha
_{k}^{n}v_{(n+k)}\}
\end{equation}
and replace eq.(\ref{DM commutator}) by
\begin{equation}
\lbrack D,M^{n,\alpha }]=nM^{n,\alpha }.  \label{Opweight}
\end{equation}

Since $M^{n,\alpha }$ acts on any definite-weight vector $v_{(k)}$ according
to
\begin{equation}
M^{n,\alpha }v_{(k)}=\alpha _{k}^{n}v_{(n+k)}  \label{M action on v}
\end{equation}
the components $(M^{n,\alpha })_{\quad j}^{i}$ of $M^{n,\alpha }$ in the
definite-weight basis are
\begin{equation}
(M^{n,\alpha })_{\quad j}^{i}=\sum_{s=-\infty }^{\infty }\alpha
_{s}^{n}\delta _{(s+n)}^{i}\delta _{j}^{(s)}  \label{generic weighted op}
\end{equation}

Now we ask for a set consisting of exactly one operator $M_{(n)}$ of each
weight. Then a single $J$-tuple of components $\alpha _{s}^{n}$ for each $%
n\in J$ characterizes the full set. We therefore drop the extraneous $\alpha
$ label, writing $M_{(n)}$ in place of $M^{n,\alpha }.$

Next, consider the Lie algebra requirement. In order for the set of
operators $M_{(n)}$ to form a Lie algebra $\mathcal{L}^{N}$ they must close
under commutation and satisfy the Jacobi identity. But the Jacobi identity
involving $M_{(m)},M_{(n),}$ and $D$ gives immediately
\begin{equation}
\lbrack D,[M_{(m)},M_{(n)}]]=(m+n)[M_{(m)},M_{(n)}]
\end{equation}
so that the commutator must be of weight $m+n,$ and therefore proportional
to $M_{(m+n)}:$%
\begin{eqnarray}
\lbrack M_{(m)},M_{(n)}] &=&\sum_{k}c_{\quad mn}^{k}M_{(k)}
\label{structure constant} \\
&=&\sum_{k}c_{mn}\delta _{m+n}^{k}\,M_{(k)}=c_{mn}M_{(m+n)}.
\end{eqnarray}
Since eq.(\ref{generic weighted op}) provides a matrix representation for
these operators (albeit doubly infinite) the Jacobi identities between $M$s
follow automatically. The commutators with Poincar\'{e} generators are
well-known once we establish that $\mathcal{L}_{N}$ is the Virasoro algebra,
so the only condition left to impose is closure under commutation. The
commutator has components
\begin{eqnarray}
([M_{(m)},M_{(n)}])_{\quad k}^{i} &=&\sum_{s=-\infty }^{\infty }(\alpha
_{s+n}^{m}\alpha _{s}^{n}-\alpha _{s}^{m}\alpha _{s+m}^{n})\delta
_{(s+m+n)}^{i}\delta _{k}^{(s)} \\
&=&c_{mn}(M_{(m+n)})_{\quad k}^{i}=c_{mn}\sum_{s=-\infty }^{\infty }\alpha
_{s}^{m+n}\delta _{(s+m+n)}^{i}\delta _{k}^{(s)}
\label{M structure constants}
\end{eqnarray}
The algebra closes if and only if, for structure constants $c_{mn}$ and all $%
s$, $m$ and $n$ the $\alpha _{s}^{n}$ satisfy the \textit{primary recursion}
relation
\begin{equation}
\alpha _{s+n}^{m}\alpha _{s}^{n}-\alpha _{s}^{m}\alpha
_{s+m}^{n}=c_{mn}\alpha _{s}^{m+n}
\end{equation}
which provides a strong constraint since the $m,n$ and $s$ dependence on the
left must factor into the product of a term depending only on $m$ and $n,$
and a term depending only on $s$ and the sum $m+n.$

We conclude this section with an aside, before completing the proof of
Theorem 1 in Sec.(4). Here, we solve the primary recursion relation for the
case of the maximally commuting, rather than the maximally non-commuting,
complete Lie algebra of definite weight operators. The calculations of
Sec.(4) are parallel to these, but require substantially more effort.

Consider a set of operators $\mathcal{J}=\{J_{n}\}$ where $J_{n}$ has
components
\begin{equation}
(J_{n})_{\quad j}^{i}=\sum_{s=-\infty }^{\infty }\alpha _{s}^{n}\delta
_{n+s}^{i}\delta _{j}^{s}
\end{equation}
in the definite-weight basis. In the maximally commuting case, $c_{mn}=0,$
and the primary recursion reduces to
\begin{equation}
\alpha _{s+n}^{m}\alpha _{s}^{n}-\alpha _{s}^{m}\alpha _{s+m}^{n}=0
\label{cmn=0 recursion}
\end{equation}
Setting $n=0$ gives $\alpha _{s}^{0}=const.$ Then, we can normalize our set
of basis vectors so that
\begin{equation}
J_{1}v_{(k)}=v_{(k+1)}
\end{equation}
i.e., $\alpha _{s}^{1}=1.$ Next, setting $n=1$ in eq.(\ref{cmn=0 recursion})
shows the constancy of all of the coefficients with a given value of $m,$%
\begin{equation}
\alpha _{s+1}^{m}=\alpha _{s}^{m}
\end{equation}
This leaves only one overall constant for each $J_{m},$ which can be
absorbed by redefining the operators. We therefore have the immediate
solution algebra, $\mathcal{J}$, of unit off-diagonal matrices
\begin{equation}
(J_{m})_{\quad j}^{i}\equiv \sum_{s=-\infty }^{\infty }\delta
_{s+m}^{i}\delta _{j}^{s}
\end{equation}
which have weight $m,$ commute with one another, and satisfy
\begin{eqnarray}
J_{m}J_{n} &=&J_{m+n} \\
J_{m}v_{n} &=&v_{m+n}
\end{eqnarray}
When the projective structure of the tower is used to allow central charges,
the algebra $\mathcal{J}$ becomes the mode algebra. The vector-valued
extension of $\mathcal{J}$ figures importantly in Sec.(8), in our discussion
of the relationships between rank $(1,2)$ tensors and string.

Notice that the action of $J_{m}$ does not distinguish among the different
vectors $v_{m},$ but has a ``spectrum'' with $\lambda _{s}=1$ for all $s$
(see eq.(\ref{M action on v})). This can be changed by adjoining to each
operator in $\mathcal{J}$ the operator $D:$%
\begin{equation}
L_{m}\equiv J_{m}D
\end{equation}
The resulting algebra, $\mathcal{J}D$ is isomorphic to the Virasoro algebra,
as is readily shown by computing
\begin{equation}
\lbrack L_{m},L_{n}]=[DJ_{m},DJ_{n}]=(n-m)J_{m+n}D=(n-m)L_{m+n}
\end{equation}
It also follows immediately that the spectrum of values $\alpha _{s}^{m}$
arising from the action of $J_{(m)}D$ on $v_{(s)}$ is nondegenerate for each
$m$.

We therefore have proved the following two results:

\begin{theorem}
The maximally commuting, complete Lie algebra of definite-weight operators
is isomorphic to the algebra $\mathcal{J}$ of unit, off-diagonal matrices.
\end{theorem}

\begin{theorem}
The algebra $\mathcal{J}D$ is isomorphic to the classical Virasoro algebra.
\end{theorem}

\smallskip

In Sec.(4) we complete the proof of Theorem 1 by carrying out steps similar
to those in the proof of Theorem 2 above, but assuming that $c_{mn}$ is
nonzero whenever $m\neq n.$ This condition is sufficient to produce the
Virasoro algebra directly.

\section{Solution of the primary recursion}

We now turn to the consequences of the primary recursion relation
\begin{equation}
\alpha _{s+n}^{m}\alpha _{s}^{n}-\alpha _{s}^{m}\alpha
_{s+m}^{n}=c_{mn}\alpha _{s}^{m+n}  \label{Primary recursion}
\end{equation}
for possible forms of $M_{(n)}.$

Before invoking eq.(\ref{Primary recursion}), we note that some of the
coefficients $\alpha _{s}^{m}$ and structure constants $c_{mn}$ are already
determined while others can be fixed by rescaling the operators $M_{(m)}$.
First, $\alpha _{s}^{m}$ and $c_{mn}$ can be related if we begin in the
adjoint representation, where from eq.(\ref{structure constant})
\begin{equation}
(M_{(k)})_{\quad j}^{i}=c_{\ kj}^{i}=\,c_{kj}\delta _{(k+j)}^{i}
\end{equation}
If eq.(\ref{generic weighted op}) is to agree with the adjoint expression we
must have
\begin{equation}
c_{kj}\delta _{(k+j)}^{i}=\sum_{s=-\infty }^{\infty }\alpha _{s}^{k}\delta
_{(s+k)}^{i}\delta _{j}^{(s)}=\alpha _{j}^{k}\delta _{(j+k)}^{i}
\end{equation}
or
\begin{equation}
c_{kj}=\alpha _{j}^{k}  \label{adjoint coeffs}
\end{equation}
from which we immediately see that $\alpha _{j}^{k}=-\alpha _{k}^{j}$ and in
particular, $\alpha _{k}^{k}=0$ for all $k.$

Next, since the unique operator of weight zero must be the dilation
generator $D,$ (since $D$ acts on weight indices only) we have
\begin{equation}
Dv_{(m)}=mv_{(m)}
\end{equation}
from which it follows that
\begin{equation}
\alpha _{m}^{0}=m
\end{equation}
while eq.(\ref{Opweight}) or eq.(\ref{adjoint coeffs}) shows that
\begin{equation}
c_{0m}=m
\end{equation}

Moving away from the adjoint representation, we now can adjust the structure
constants $c_{-1,m}$ by rescaling each $M_{(m)}$ by a factor $\beta _{m}.$
Such rescaling does not alter $\alpha _{m}^{m}=0.$ Then from eq.(\ref{M
structure constants})
\begin{equation}
c_{-1,m}\equiv c_{-1,m}^{new}=\frac{\beta _{m-1}}{\beta _{-1}\beta _{m}}%
\,c_{-1,m}^{old}
\end{equation}
Because $c_{mn}$ is antisymmetric, $c_{-1,-1}$ vanishes. Choosing the $\beta
s$ so that
\begin{eqnarray}
\beta _{m} &=&\frac{c_{-1,m}^{old}}{(m+1)\beta _{-1}}\beta _{m-1}\qquad
\forall m\geq 0  \label{beta+} \\
\beta _{m-1} &=&(m+1)\frac{\beta _{-1}}{c_{-1,m}^{old}}\beta _{m}\qquad
\forall m\leq -2  \label{beta-}
\end{eqnarray}
fixes the remaining $c_{-1,m}$ so that
\begin{equation}
c_{-1,m}=m+1=-c_{m,-1}\qquad \forall m
\end{equation}
Note that since the commutator of two distinct operators never vanishes, we
cannot have $c_{-1,m}=0$ for $m\neq -1$. Eqs.(\ref{beta+}) and (\ref{beta-})
fix all but two of the $\beta _{m}$: we still may freely fix any one
coefficient in the set $\left\{ \beta _{-1},\beta _{1},\beta _{2},\beta
_{3}\ldots \right\} $ and any second coefficient in the set $\left\{ \beta
_{-2},\beta _{-3},\beta _{-4}\ldots \right\} $.

Finally, we can redefine the basis $v_{(m)}$ in terms of the action of
either $M_{(1)}$ or $M_{(-1)}.$ That is, it is always possible to replace $%
v_{(m)}$ by $\eta _{m}v_{(m)}$ and choose the constants $\eta _{m}$ so that
\begin{equation}
M_{(-1)}v_{(m)}=(m+1)v_{(m-1)}  \label{vecnorm}
\end{equation}
This normalization of $v_{(m)}$ again scales the values of the $\alpha
_{s}^{m}$ in eq.(\ref{generic weighted op}), but has no other effect. The
values of $\alpha _{m}^{-1}$ given by eq.(\ref{vecnorm})
\begin{equation}
\alpha _{m}^{-1}=m+1  \label{alpha-1,m}
\end{equation}
are consistent with eq.(\ref{Primary recursion}).

These normalizations together with eq.(\ref{Primary recursion}) are
sufficient to determine all of the remaining $\alpha _{s}^{m}.$ The proof
begins by setting $n=-1$ and $m=1$ in eq.(\ref{Primary recursion}) (since
the equation is an identity if either $m=0$ or $n=0$). Using eq.(\ref
{alpha-1,m}), this gives
\begin{equation}
(s+1)\alpha _{s-1}^{1}-(s+2)\alpha _{s}^{1}=-2s
\end{equation}
and induction readily yields $\alpha _{s}^{1}=s-1$ for all $s\neq -2.$ The
value $\alpha _{-2}^{1}=-3$ may be fixed by the choice of $\beta _{1},$ so
we have
\begin{equation}
\alpha _{s}^{1}=s-1,\qquad \forall s
\end{equation}

Collecting our results so far, we have
\begin{eqnarray}
\alpha _{s}^{1} &=&s-1 \\
\alpha _{s}^{0} &=&s,\qquad \qquad c_{0s}=s \\
\alpha _{s}^{-1} &=&s+1,\qquad c_{-1,s}=s+1
\end{eqnarray}
for all $s$, where we are still free to fix any one element of the set $%
\left\{ \beta _{-2}, \beta _{-3}, \beta _{-4}\ldots \right\} $.

Since $M_{(m)}$ with $m\in \{-1,0,1\}$ form a closed subalgebra, we do not
yet have sufficient seed values to use eq.(\ref{Primary recursion})
recursively. Therefore, we next consider eq.(\ref{Primary recursion}) for
three cases: $(n=-1,m=2),(n=1,m=-2)$ and $(n=2,m=-2).$ The first gives
\begin{equation}
(s+3)\alpha _{s}^{2}-(s+1)\alpha _{s-1}^{2}=3(s-1)
\end{equation}
which determines
\begin{eqnarray}
\alpha _{s}^{2} &=&s-2\qquad \forall s\neq -2,-3 \\
\alpha _{-2}^{2}+\alpha _{-3}^{2} &=&-9
\end{eqnarray}
The second case is similar except that $c_{-2,1}$ is unknown:
\begin{equation}
(s-1)\alpha _{s+1}^{-2}-(s-3)\alpha _{s}^{-2}=(s+1)c_{-2,1}
\label{(n=1,m=-2) recursion}
\end{equation}
Eq.(\ref{(n=1,m=-2) recursion}) permits $\alpha _{s}^{-2}$ to be found in
terms of $c_{-2,1}$ for all $s<2$ and $s>3,$ and provides one remaining
relation
\begin{equation}
\alpha _{3}^{-2}+\alpha _{2}^{-2}=3c_{-2,1}
\end{equation}
Finally, the $(n=2,m=-2)$ case,
\begin{equation}
\alpha _{s+2}^{-2}\alpha _{s}^{2}-\alpha _{s}^{-2}\alpha _{s-2}^{2}=sc_{-2,2}
\end{equation}
has useful special cases. For $s=4$ we find $\alpha _{6}^{-2}=2c_{-2,2}$
while for $s=2$ we have $\alpha _{2}^{-2}=c_{-2,2}.$ From the expression for
$\alpha _{6}^{-2}$ in terms of $c_{-2,1}$ already determined from eq.(\ref
{(n=1,m=-2) recursion}) it follows that
\begin{equation}
4c_{-2,1}=3c_{-2,2}
\end{equation}
Therefore, using $\beta _{-2}$ to set $\alpha _{2}^{-2}=4$ immediately
yields $c_{-2,2}=4$ and $c_{-2,1}=3.$ With these structure constants
determined, the remaining values, $\alpha _{s}^{\pm 2}=s\mp 2$
follow quickly for all $s$.

Now two induction arguments give the remaining $\alpha _{s}^{m}$. For $m<-2$
we use $n=-1$ in eq.(\ref{Primary recursion}) with $\alpha _{s}^{-2}=s+2$ as
the seed value for the induction. Substituting for known values and
rearranging
\begin{equation}
\alpha _{s}^{m-1}=\frac{1}{(m+1)}\left( (s+m+1)\alpha _{s}^{m}-(s+1)\alpha
_{s-1}^{m}\right)
\end{equation}
gives the expected values $\alpha _{s}^{m}=s-m$ for all $m<-2$ and all $s.$

To accomplish the same result for positive $m$ requires eq.(\ref{Primary
recursion}) with $n=1,$ but now we also need values for $c_{1m}$. There are
several steps needed to find $c_{1m}.$ First set $n=-1$ and $s=-1$ in eq.(%
\ref{Primary recursion}) to find $\alpha _{-1}^{m}$ for positive $m.$ The
resulting values, $\alpha _{-1}^{m}=-(m+1),$ together with the $(n=-1,s=0)$
equation give $\alpha _{0}^{m}=-m.$ Then setting $n=-1$ and $s=1$ yields $%
\alpha _{1}^{m}=1-m.$ Finally, using these values of $\alpha _{0}^{m}$ and $%
\alpha _{1}^{m}$ in eq.(\ref{Primary recursion}) with $n=+1$ produces the
necessary result, $c_{m1}=1-m.$ With these values for the structure
constants, the $n=+1$ equation becomes
\begin{equation}
\alpha _{s}^{m+1}=\frac{1}{(m-1)}\left( (s+m-1)\alpha _{s}^{m}-(s-1)\alpha
_{s+1}^{m}\right)   \label{(+1) recursion}
\end{equation}
and induction based on $\alpha _{s}^{2}=s-2$ works as expected.

Combining everything, we have simply
\begin{equation}
\alpha _{s}^{m}=s-m=c_{ms},\qquad \forall m,\forall s
\end{equation}
completing the proof that conditions $(1-4)$ imply the Virasoro form for the
operators $M_{(m)}.$ This concludes the proof of Theorem 1.

\smallskip

Before moving on to alternate representations for weight maps in the next
section, we note that the commuting algebra $\mathcal{J}$ makes it possible
to shift $\alpha _{s}^{m}$ by an arbitrary amount without changing the
Virasoro commutators. Let
\begin{equation}
M_{(k)}^{\alpha }=M_{(k)}+\alpha J_{(k)}
\end{equation}
so that
\begin{equation}
(M_{(k)}^{\alpha })_{\quad j}^{i}=\sum_{s=-\infty }^{\infty }(s-m+\alpha
)\delta _{s+m}^{i}\delta _{j}^{s}
\end{equation}
for an arbitrary constant $\alpha .$ It is straightforward to check that $%
M_{(k)}^{\alpha }$ satisfies
\begin{equation}
\lbrack M_{(k)}^{\alpha },M_{(l)}^{\alpha }]=(l-k)M_{(k+l)}^{\alpha }
\end{equation}
for any $\alpha .$ In particular, the choice $\alpha =m$ simplifies the
representation to the form
\begin{equation}
(M_{(k)}^{\alpha =m})_{\quad j}^{i}=\sum_{s=-\infty }^{\infty }s\delta
_{s+m}^{i}\delta _{j}^{s}
\end{equation}
which we will use in Sec.(5).

\section{Alternative representations of weight maps}

Many calculations on the infinite tower of weight states are greatly
simplified by the use of a continuous basis, which we define below.
Essentially, we use the infinite dimensional $(0,1)$ vectors as the
coefficients in a Fourier series. Then a matrix operator on the $(0,1)$
vector space may be written as a double Fourier series. The operator becomes
a function of two variables, a fact which lets us represent vector-valued
operators as strings in Sec.(7). The product of two operators is given by
integration over the ``inner'' pair of variables while the restriction to
definite weight operators leads ultimately to wave solutions. In this
section we develop these ideas, keeping as our primary motivation the
representation of scale-invariant operators in $\mathcal{L}^{N}$.

Given what we already know, we can find motivation for moving to a Fourier
representation in the shifted sum between $\alpha _{s+n}^{m}$ and $\beta
_{s}^{n}$ of the multiplication rule
\begin{equation}
(M_{(m)}N_{(n)})_{\quad j}^{i}=\sum_{s=-\infty }^{\infty }\alpha
_{s+n}^{m}\beta _{s}^{n}\delta _{(s+m+n)}^{i}\delta _{(s)j}
\label{convolution}
\end{equation}
Noting that a similar shifted sum occurs when multiplying Fourier
transforms, we can try to reformulate the multiplication rule in terms of
Fourier series. It turns out that the simplest anzatz works: we can model
the maps by treating their components as the coefficients of a double
Fourier series, that is, we define a distribution of two variables in terms
of the components, $(M)_{\quad n}^{m}$ of an arbitrary operator. However,
having noted this clue that a Fourier representation will work, it is most
logical to begin the formal development with the continuous representation
of $(0,1)$ vectors and build up from there.

Therefore, we start with the definition of the function space on which the
Virasoro operators act, using the components of $(0,1)$ vectors in the
definite weight basis as coefficients in a Fourier series. Beginning with
the definite weight basis $(v_{\left( k\right) })^{m}=\delta _{k}^{m}$ we
find that the basis functions
\begin{equation}
f_{k}(x)=\sum_{m=-\infty }^{\infty }(v_{k})^{m}e^{imx}=e^{ikx}\,  \label{fk}
\end{equation}
where $x\in [-\pi ,\pi ],$ are individual Fourier modes.\footnote{%
Certain choices have been made in writing eq.(\ref{fk}). We could equally
well take the real part of this and subsequent expressions so that $%
f:[-\pi,\pi ]\rightarrow \mathcal{R}$. Also, notice that the choice of the
region $[-\pi ,\pi ]$ is arbitrary. Not only could we pick any other bounded
interval, but we could use a circle instead so that $f:\mathcal{S}
^{1}\rightarrow \mathcal{R}.$ The only strict requirement is the use of a
compact region, since compactness guarantees the existence of Fourier series
for piecewise continuous functions. Even this requirement may be relaxed if
we allow the index set to be $R$ rather than $J,$ permitting the use of a
Fourier integral representation on the real line (and consequently, an
``infinite length string'' representation). It is important to remember as
we look at different representations that the only required property is the
representation of weight maps.} For a general linear combination\footnote{%
Recall that indefinite weight combinations cannot be measured directly. For
a crude example, it is as if we were to make a vector of position, momentum
and force, $v^{\mu }=(\ldots ,v_{-2}^{\mu },v_{-1}^{\mu },v_{0}^{\mu
},v_{1}^{\mu },\ldots )=(\ldots ,F^{\mu },p^{\mu },0,x^{\mu },\ldots )$ and
look at the dynamics of the entire object at once. Nonetheless, as long as
we restrict any physical predictions to states of zero weight, there is no
reason indefinite vectors $v$ cannot be used. In particular, notice that the
whole of $v^{\mu }$ transforms as a Lorentz vector, consistent with the
decoupling of Lorentz and weight transformations.} of the basis vectors
\begin{equation}
(v)^{m}=\sum_{k=-\infty }^{\infty }\alpha ^{k}(v_{k})^{m}
\end{equation}
we can write the Fourier series
\begin{equation}
f(x)=\sum_{m=-\infty }^{\infty }\sum_{k=-\infty }^{\infty }\alpha
^{k}(v_{k})^{m}e^{imx}=\sum_{k=-\infty }^{\infty }\alpha ^{k}e^{ikx}
\end{equation}
but it is clear that we need (in either representation) some convergence
criterion. Such a criterion is provided most easily by a norm.

Choosing a suitable norm for the functions $f(x)$ is not as simple as it
seems. Setting $g(x)=\sum_{k=-\infty }^{\infty }\beta ^{k}e^{ikx}$ and
allowing the coefficients $\alpha ^{k},\beta ^{k}$ to be complex (as they
are, for example, if $f(x^{\mu };x)$ and $g(x^{\mu };x)$ below are complex
scalar fields on spacetime) the usual Hilbert space norm,
\begin{equation}
(f,g)\equiv \frac{1}{2\pi }\int\limits_{-\pi }^{\pi }f\,^{*}(x)g(x)dx=%
\sum_{k=-\infty }^{\infty }(\alpha ^{k})^{*}\beta ^{k}
\end{equation}
is unstable under conformal transformations. To see this, note that under a
scale change $e^{\varphi },$ the function $f(x)$ becomes
\begin{equation}
f^{\prime }(x)=\sum_{k=-\infty }^{\infty }\alpha ^{k}e^{k\varphi }e^{ikx}
\end{equation}
so that
\begin{equation}
(f^{\prime },f^{\prime })=\sum_{k=-\infty }^{\infty }(\alpha ^{k})^{*}\alpha
^{k}e^{2k\varphi }
\end{equation}
which is only guaranteed to converge for series $\{\alpha ^{k}\}$ having
greater than exponential convergence. This constrains the function space
quite severely (although it is consistent with the Gaussian envelopes of
typical wave packets). It is more natural and satisfactory to begin with a
scale invariant norm.

To implement scale invariance of the norm we first introduce the \textit{%
conformal conjugate} of a vector $g(x)=\sum_{k=-\infty }^{\infty }\beta
^{k}e^{ikx}$ by
\begin{equation}
\bar{g}(x)\equiv \sum_{k=-\infty }^{\infty }\beta ^{-k}e^{ikx}
\end{equation}
It is convenient to used raised and lowered indices in the usual way,
writing $\beta ^{k}\rightarrow \beta _{k}\equiv \beta ^{-k}.$ Then a
conformally invariant inner product may be written using the summation
convention
\begin{equation}
f\cdot g=\alpha ^{m}\beta _{m}=\alpha _{m}\beta ^{m}
\end{equation}
This inner product is the natural weight tower extension of the Killing
metric of the conformal group. When the coefficients $\beta _{k}$ are real,
conformal and complex conjugation are equivalent
\begin{equation}
\bar{g}(x)=\sum_{k=-\infty }^{\infty }\beta _{k}e^{ikx}=\sum_{k=-\infty
}^{\infty }\beta ^{-k}e^{ikx}=\sum_{m=-\infty }^{\infty }\beta
^{m}e^{-imx}=g^{*}(x)
\end{equation}

Returning to the norm, we impose \textit{both} complex and conformal
conjugation.
\begin{equation}
\langle f,g\rangle \equiv \frac{1}{2\pi }\int\limits_{-\pi }^{\pi }\bar{f}
\,^{*}(x)g(x)dx=\sum_{m=-\infty }^{\infty }\alpha _{m}^{*}\beta ^{m}=\langle
g,f\rangle ^{*}  \label{conformal inner product}
\end{equation}
This norm is real but not positive definite, since
\begin{eqnarray}
\langle f,f\rangle &=&\sum_{m=-\infty }^{\infty }\alpha _{m}^{*}\alpha ^{m}
\\
&=&\alpha _{0}^{*}\alpha _{0}+\sum_{m=1}^{\infty }[\alpha ^{-m*}\alpha
^{m}+\alpha ^{m}{}^{*}\alpha ^{-m}]
\end{eqnarray}
is linear, not quadratic, in $\alpha ^{m}.$ However, there are many regions
of the space of functions $f(x)$ for which this expression has definite
sign. Most of these depend on careful matching of phases, but two classes
are easier to describe and a third deserves mention.

Consider the class of functions $f$ with Fourier coefficients in the
definite weight basis satisfying $\alpha _{k}=\lambda _{k}\alpha ^{k}$ for
all $k>0$ and arbitrary numbers $\lambda _{k}>0.$ For such functions (which
are gauge-equivalent to the class of self-dual functions, $\bar{f}(x)=f(x),$
identical to the symmetric functions), the norm $\langle f,g\rangle $ is
\begin{equation}
\langle f,f\rangle =\sum_{m=-\infty }^{\infty }\lambda _{m}|\alpha ^{m}|^{2}
\end{equation}
which is positive definite and vanishes if and only if $f=0.$ Unfortunately
the presence of $\lambda _{k}$ again destablilzes the class, since only
certain sequences $\{\lambda _{k}\}$ give convergence. We might consider
restricting $\lambda _{k}\leq 1,$ but then a function $f$ with $\alpha
_{k}=\lambda _{k}\alpha ^{k}$ will have a conformal dual with $\alpha _{k}=%
\frac{1}{\lambda _{k}}\alpha ^{k},$ and therefore not in the space unless $%
\lambda _{k}=1.$ We are therefore constrained to the self dual functions.
Convergence in the self-dual norm gives a Hilbert space, $\mathcal{H}$. This
result is of considerable importance, since it shows that \textit{there is a
Hilbert space of conformally self-dual vectors automatically associated with
the tangent space at each point of any scale-invariant geometry}.

A second class of functions, those which are anti-self-dual, are similarly
shown to form a space with negative definite norm.

A third class of functions of interest are those for which $\alpha _{k}=0$
for all $k\geq 0.$ Any such function $f$ has only positive frequency modes
in its Fourier expansion
\begin{equation}
f(x)=\sum_{k=1}^{\infty }\alpha ^{k}e^{ikx}
\end{equation}
Consequently, all are null, $\langle f,f\rangle =0$ since the conformal
conjugate vanishes. Of course, the presence of null vectors prevents our
using this norm to define a Hilbert space for these functions, but now the
usual Hilbert norm $(f,g)$ is stable for \textit{contractive }gauge changes.

\smallskip

We now move from the continuous representation of vectors to consider the
continuous representation of weight maps. Writing maps as double Fourier
series with components in the definite weight basis, $f_{k}(x),$ the maps
become functions of two variables, $(x,y)\in \mathcal{N}\equiv [-\pi ,\pi ]%
\times [-\pi ,\pi ].$ Since the weight maps $M$ of previous sections take $%
(0,1)$ vectors to $(0,1)$ vectors, the inner product requires one index in
each position $(M)_{\quad n}^{m}$. A generic weight map therefore takes the
form
\begin{equation}
M(x,y)\equiv \sum_{m,n}(M)_{\quad n}^{m}e^{imx}e^{-iny}  \label{mixed map}
\end{equation}

The product of two operators follows from eq.(\ref{conformal inner product})
as
\begin{eqnarray}
MN &\equiv &\frac{1}{2\pi }\int\limits_{-\pi }^{\pi }M(x,z)N(z,y)dz
\label{generic op product} \\
&=&(2\pi )^{-1}\int \sum_{m,n}(M)_{\quad n}^{m}\sum_{p,q}(N)_{\quad
q}^{p}\exp imx\exp -i(n-p)z\exp -iqy \\
&=&\sum_{m,q}(\sum_{n}(M)_{\quad n}^{m}(N)_{\quad q}^{n})\exp imx\exp -iqy \\
&=&(MN)(x,y)
\end{eqnarray}
which shows that the sign convention of eq.(\ref{mixed map}) and the
definition of the product via eq.(\ref{generic op product}) are in agreement
with the matrix representation.

It is illustrative to use the $2$-dim representation to find the commutator
of a general operator with $D,$ and to rederive the generic form, eq.(\ref
{generic weighted op}), of a definite weight operator. The convention for
indices introduced above (which ties together complex and conformal
conjugation) requires us to write
\begin{eqnarray}
D(x,y) &=&(2\pi )^{-1}\sum_{m,n}(D)_{\quad n}^{m}e^{imx}e^{-iny} \\
&=&(2\pi )^{-1}\sum_{m,n}\sum_{s}s\delta _{s}^{m}\delta
_{n}^{s}e^{imx}e^{-iny} \\
&=&(2\pi )^{-1}\sum_{s}se^{is(x-y)} \\
&=&-i\frac{\partial }{\partial x}\delta (x-y)  \label{D in def wt basis}
\end{eqnarray}
We now show that this simple form for $D(x,y)$ leads to a set of definite
weight functions $f_{k}(x)$ with only pure modes $f_{k}(x)\sim e^{ikx}$, and
to a set of definite weight operators $M_{(k)}(x,y)$ with only ``right
moving modes'', $M_{(k)}(x,y)\sim M_{(k)}(x-y).$ While some such restriction
on operators is a necessary concomitant of definite-weight tensors,
``left-moving modes'' are present in all but the definite-weight basis. In
cases where both left- and right-moving modes appear, they are not
independent but are determined by a single analytic function. We show below
that operators with truly independent left- and right-moving modes can be
constructed using operators of definite \textit{anti}-weight.

To begin, consider the action of $D(x,y)$ on a $(0,1)$ vector
\begin{equation}
f(x)=\sum\limits_{k=-\infty }^{\infty }c^{k}e^{ikx}
\end{equation}
Note that regardless of the values of the $c^{k}$ we have $f(\pi )=f(-\pi ).$
The condition for $f(x)$ to be of definite weight is
\begin{equation}
Df=\lambda f  \label{Eigenvalue eq.}
\end{equation}
or, in terms of eq.(\ref{generic op product})
\begin{equation}
Df\equiv \int\limits_{-\pi }^{\pi }D(x,y)f(y)dy  \label{Df}
\end{equation}
Unfortunately, the integration by parts used to evaluate eq.(\ref{Df}) gives
a surface term of the form
\begin{equation}
i\delta (x-\pi )f(\pi )-i\delta (x+\pi )f(-\pi )  \label{Left surface term}
\end{equation}
which vanishes only if $f(\pi )=f(-\pi )=0.$ In combination with the
eigenvalue relation, eq.(\ref{Eigenvalue eq.}), $f(\pm \pi )=0$
is inconsistent.

Therefore, in order to get a complete set of weighted functions satisfying
eq.(\ref{Eigenvalue eq.}) we must include a singular correction term in the
definition of $D$. The most general term which can be used to cancel terms
of the form of eq.(\ref{Left surface term}) is
\begin{eqnarray}
c(x,y) &=&c_{1}\delta (x+\pi )\delta (y+\pi )+c_{2}\delta (x+\pi )\delta
(y-\pi )  \nonumber \\
&&+c_{3}\delta (x-\pi )\delta (y+\pi )+c_{4}\delta (x-\pi )\delta (y-\pi )
\label{c(x,y)}
\end{eqnarray}
We now show that the distribution $c(x,y)$ can be chosen to eliminate all
surface terms, whether we apply $D$ from the right or from the left.
Moreover, the choice for $c(x,y)$ works the same way for the action of $D$
on any rank of tensor.

To find $c(x,y),$ let $D$ be given by
\begin{equation}
D(x,y)=-i\frac{\partial }{\partial x}\delta (x-y)-ic(x,y)
\end{equation}
with $c(x,y)$ given by eq.(\ref{c(x,y)}), and consider the right and left
action of $D$ on the $r^{th}$ slot of a rank $(0,k)$ tensor, that is,
\begin{equation}
(DT)_{r}\equiv \int\limits_{-\pi }^{\pi }dzD(x,z)T(x_{1,}x_{2},\ldots
,x_{r-1},z,x_{r-2},\ldots ,x_{k})
\end{equation}
and
\begin{equation}
(TD)_{r}\equiv \int\limits_{-\pi }^{\pi }dzT(x_{1,}x_{2},\ldots
,x_{r-1},z,x_{r-2},\ldots ,x_{k})D(z,x)
\end{equation}
where
\begin{equation}
T(x_{1,}x_{2},\ldots ,z,\ldots ,x_{k})=\sum T^{n_{1}\ldots n\ldots
n_{k}}e^{in_{1}x_{1}+\ldots +inz+\ldots +in_{k}x_{k}}
\end{equation}
For $(DT)_{r}$ the integration by parts gives a surface term of the form
\begin{eqnarray}
S.T. &=&-i\left[ -\delta (x-z)T(x_{1,}x_{2},\ldots ,z,\ldots ,x_{k})\right]
_{-\pi }^{\pi }  \nonumber \\
&&-i\int\limits_{-\pi }^{\pi }c(x,z)T(x_{1,}x_{2},\ldots ,z,\ldots ,x_{k})
\end{eqnarray}
which cancels provided
\begin{eqnarray}
c_{1}+c_{2} &=&1 \\
c_{3}+c_{4} &=&-1
\end{eqnarray}
The surface term for $(TD)_{r}$ also vanishes, provided
\begin{eqnarray}
c_{1}+c_{3} &=&-1 \\
c_{2}+c_{4} &=&1
\end{eqnarray}
Both conditions are satisfied if $D$ is of the form
\begin{equation}
D(x,y)=-i[\partial _{x}\delta (x-y)-\delta (x-\pi )\delta (y+\pi )+\delta (x+%
\pi )\delta (y-\pi )]  \label{D(x,y) with S.T.}
\end{equation}
plus an arbitrary multiple of the purely surface term
\begin{eqnarray}
E &=&[\delta (x-\pi )\delta (y-\pi )-\delta (x-\pi )\delta (y+\pi ) \\
&&-\delta (x+\pi )\delta (y-\pi )+\delta (x+\pi )\delta (y+\pi )]
\end{eqnarray}
Since $ET=TE=0$ for any tensor $T$ (as long as $T(\pi )=T(-\pi )$) we can
simply drop the $E$ term.

Now that the surface terms vanish, the left and right action of $D(x,y)$ on
the $r^{th}$ slot of any tensor is simply given by the integrated part,
\footnote{%
In evaluating $\int f(y)\delta (x-y)$ we have taken the full value of the $%
\delta$-functions at the endpoints $\pm \pi .$ Other choices can
be made by adjusting the surface term.}
\begin{eqnarray}
(DT)_{r} &=&-i\frac{\partial T}{\partial x_{r}}  \label{DT} \\
(TD)_{r} &=&i\frac{\partial T}{\partial x_{r}}  \label{TD}
\end{eqnarray}
Using eqs.(\ref{DT}) and (\ref{TD}) we now find a complete set of
eigenfuntions and definite weight operators.

Substituting $D(x,y)$ into eq.(\ref{Eigenvalue eq.}), we find the
differential equation
\begin{equation}
Df^{(k)}=-i\frac{\partial f^{(k)}(x)}{\partial x}=kf^{(k)}(x)  \label{Df(x)}
\end{equation}
which is solved by the set of pure mode functions
\begin{equation}
f^{(k)}(x)=A_{k}e^{ikx}
\end{equation}
Thus, just like plane waves in quantum mechanics, the simplest set of
eigenfunctions of $D(x,y)$ do not lie in the Hilbert space. It is possible
to develop a self-dual basis, although $D(x,y)$ then takes a more
complicated form.

For rank two tensors (or equivalently rank $(r,2)$ for any $r$) we can
compute the commutator,
\begin{equation}
\lbrack D,M](x,y)=(DM)_{1}-(MD)_{2}
\end{equation}
which allows us to define operators $M_{(k)}$ of definite weight by
demanding
\begin{equation}
\lbrack D,M_{(k)}]=kM_{(k)}
\end{equation}
Substituting from eqs.(\ref{DT}) and (\ref{TD}),
\begin{equation}
\left( \frac{\partial }{\partial x}+\frac{\partial }{\partial y}\right)
M(x,y)=ikM(x,y)
\end{equation}
with the immediate solution:
\begin{equation}
M_{(k)}(x,y)=e^{ikx}\,h_{k}(x-y)=e^{ikx}\,\sum_{m=-\infty }^{\infty }\alpha
_{m}^{k}e^{im(x-y)}  \label{M(x,y) of def weight}
\end{equation}
where $h$ is an arbitrary function, which we take to be analytic. Notice
that $M_{(k)}(x,y)$ has only ``right-moving'' modes.

It is natural enough to wonder whether varying the treatment of the
continuous representation above could lead to ``left-moving'' modes as well,
in the expression for $M_{(k)}(x,y).$ As mentioned at the start of the
Section, it is indeed possible to choose a different basis than the pure
mode $f_{(k)}(x)$ used here, such that the result for $M_{(k)}(x,y)$ is a
superposition of right and left modes. As one might expect, the amplitudes
of the two modes are still expressed in terms of a \textit{single} function $%
h(z),$ rather than two independent functions $f_{1}(x+y)$ and $f_{2}(x-y).$
The full expressions for $D,M_{(k)},$ and $f_{(k)}$ in a set of alternative
bases are derived in Appendix 1.

A superposition of \textit{independent} left and right moving modes arises
only if we consider operators which have definite weight under\ \textit{anti}
-commutation with the dilation generator,
\begin{equation}
\{D,N_{(k)}\}\equiv DN_{(k)}+N_{(k)}D=kN_{(k)}  \label{Anticommutation}
\end{equation}
Eqs.(\ref{DT}) and (\ref{TD}) lead immediatly to
\begin{equation}
\left( \frac{\partial }{\partial x}-\frac{\partial }{\partial y}\right)
N_{(k)}(x,y)=ikN_{(k)}(x,y)
\end{equation}
with the solution
\begin{equation}
N_{(k)}(x,y)=e^{ikx}\,g_{k}(x+y)
\end{equation}
Any operator satisfying eq.(\ref{Anticommutation}) will be said to have
\textit{anti-weight} $k.$ Notice that the distinct behavior of
``right-moving'' and ``left-moving'' modes displayed by $M_{(k)}(x,y)$ and $%
N_{(k)}(x,y)$ is a form of heterosis (see [1], pg 306). We stress that this
heterosis is distinct from that of the heterotic string. For string, the
asymmetry is due to the assignment of bosonic and fermionic operator types
to the left and right moving mode amplitudes, and does not result in
anticommutation with the dilation operator.

The operators $N_{(k),}$ which anticommute with $D,$ are just as useful for
constructing zero-weight action functionals as the $M_{(k)}$ because such
action functionals will generally be quadratic (or of even powers) in
operator-valued fields. Thus, in a term in an action such as
\begin{equation}
S\sim \int N_{(k)}N_{(k)}
\end{equation}
the quadratic expression $N_{(k)}N_{(k)}$ commutes with $D$ just as the
corresponding expression $M_{(k)}M_{(-k)}$ does.

It is easy to check using anticommutation with $D,$ that the products $%
M_{m}N_{n}$ and $N_{n}M_{m}$ have weights $n+m$ and $n-m,$ respectively.
Interestingly, the products $N_{m}N_{n}$ and $N_{n}N_{m}$ also have
different weights, $m-n$ and $n-m.$ This leads to complicated commutation
relations for the $N_{m}.$ However, the use of t'Hooft commutators restores
simplicity. Working with,
\begin{equation}
N_{m}(x,y)=e^{ikx}\,\sum_{m=-\infty }^{\infty }me^{im(x+y)}
\end{equation}
we define the t'Hooft commutator as
\begin{equation}
\lbrack N_{m},N_{n}]_{tH}\equiv N_{m}N_{n}-e^{2i(m-n)x}N_{n}N_{m}
\end{equation}
Then the extra phase equalizes the weights and we find
\begin{equation}
\lbrack N_{m},N_{n}]_{tH}=(m-n)M_{m-n}
\end{equation}
Similarly, the t'Hooft commutator of $M_{m}$ with $N_{n}$ is proportional to
$N_{m+n},$ so the full set $\{M_{m},N_{n}\}$ form a quantum group with
modular symmetry.

The most general linear combination of definite weight operators, including
both commuting and anticommuting types, is
\begin{eqnarray}
M(x,y) &=&\sum_{k=-\infty }^{\infty }\left[ \lambda _{k}M_{(k)}(x,y)+\eta
_{k}N_{(k)}(x,y)\right] \\
&=&\sum_{k=-\infty }^{\infty }e^{ikx}\,\sum_{m=-\infty }^{\infty }\left[
\alpha _{m}^{k}e^{im(x-y)}+\beta _{m}^{k}e^{-im(x+y)}\right]
\label{Gen def we M(x,y)}
\end{eqnarray}
where some of the operator properties, including the commutation and
anticommutation with $D,$ are contained in the $(x,y)$-dependence of $M(x,y).
$ This is an important notational difference from standard quantum field
theory expressions, in which the \textit{amplitudes} are the operators.
Here, the amplitudes are simply numbers. In fact, the operator content of
these expressions is distributed, because it is the amplitudes that govern
\textit{which} algebra the operators represent. For example, in the
expression above we might choose $\alpha _{m}^{k}=m,\beta _{m}^{k}=0$ so
that $M(x,y)$ lies in the Virasoro algebra, $\mathcal{L}^{N},$ or we could
set $\alpha _{m}^{k}=1,\beta _{m}^{k}=0$ so that $M(x,y)$ is an element of
the mode algebra, $\mathcal{J}$. This distributed character of the operator
properties complicates the comparison with standard results.

The matrix results of the preceeding sections may also be found using the
continuous representation. Substitution of the Fourier series for $%
M_{(k)}(x,y)$ shows the required agreement with eq.(\ref{generic weighted op}%
). It is also straightforward to verify the convolution form of eq.(\ref
{convolution}) for the product of two operators, as well as the condition
for closure under commutation, eq.(\ref{Primary recursion}).

In the specific case of the Virasoro operators, for which we can take $%
\alpha _{s}^{k}=s,$ the distributional character of the Virasoro operators
in $\mathcal{L}_{N}$ may be expressed directly:
\begin{eqnarray}
M_{(k)}(x,y) &=&(2\pi )^{-1}\sum_{m,n}\sum_{s=-\infty }^{\infty }s\delta
_{s+k}^{m}\delta _{n}^{s}e^{imx}e^{-iny} \\
&=&(2\pi )^{-1}\sum_{n}ne^{ikx}e^{in(x-y)}  \nonumber \\
&=&(2\pi )^{-1}e^{ikx}(-i\frac{\partial }{\partial x})\sum_{n}e^{in(x-y)}
\nonumber \\
&=&-ie^{ikx}\frac{\partial }{\partial x}\delta (x-y)  \nonumber
\end{eqnarray}
where we have inserted an overall factor of $(2\pi )^{-1}$ in the
normalization for convenience. Unfortunately, this distributional form for $%
M_{(k)}$ leads to surface terms when $M_{(k)}$ acts on arbitrary tensor
fields. While it may be possible to find corrective additional terms as we
did for $D(x,y),$ it is simpler to eliminate the surface terms by going to a
derivative representation. To find the derivative representation, we easily
check that the integral action of $M_{(k)}(x,y)$ on the $m$th weight slot of
an arbitrary rank $(r,s)$ tensor $T^{\mu \cdots \nu }(x_{1},\ldots x_{s})$
may be written as
\begin{equation}
M_{(k)}T=\int dzM_{(k)}(x,z)T(\cdots ,z,\cdots )=L_{k}(x)T(\cdots ,x,\cdots
)+surface\ terms  \label{M to L}
\end{equation}
where $L_{k}$ is the\textit{\ }operator
\begin{equation}
L_{k}\equiv -ie^{ikx}\frac{\partial }{\partial x}
\end{equation}
The generators $L_{k}$ are immediately recognizable as the generators of the
diffeomorphism group of the projective punctured plane, or equivalently, the
diffeomorphisms of $S^{1},$ provided we interpret $x$ as an angular
variable. It is easy to check that
\begin{equation}
\left[ L_{m},L_{n}\right] =(n-m)L_{m+n}
\end{equation}

The action of $L_{k}$ on a vector $f(x)$ is
\begin{eqnarray}
L_{k}f(x) &=&-ie^{ikx}\frac{\partial }{\partial x}\quad v\sum_{m=-\infty
}^{\infty }\beta _{m}e^{imx} \\
&=&v\sum_{m=-\infty }^{\infty }m\beta _{m}e^{i(k+m)x}  \nonumber
\end{eqnarray}
and when $f$ reduces to a particular mode $f_{m}$
\begin{equation}
L_{k}f_{m}(x)=mv\exp i(k+m)x=mf_{m+k}
\end{equation}
as expected.

These calculations may be repeated for the commuting algebra $\mathcal{J}$.
As noted following eq.(\ref{Gen def we M(x,y)}), it is only the choice of $%
\alpha _{s}^{n}$ in eq.(\ref{Gen def we M(x,y)}) that fixes the algebra to
be $\mathcal{J}$ instead of $\mathcal{L}^{N}.$ We immediately see that $%
\mathcal{J}$ may be represented by the operators
\begin{equation}
J_{n}=e^{inx}\delta (x-y)
\end{equation}
with the identity operator given by $J_{0}=\delta (x-y).$ Then the product
of two $J$s is simply
\begin{equation}
J_{n}J_{m}=J_{n+m}
\end{equation}
so that
\begin{equation}
\lbrack J_{n},J_{m}]=0
\end{equation}
By analogy with the definition of $L_{k}$ we can define operators $H_{m}$
satisfying
\begin{eqnarray}
(JT)(x_{1},\cdots ,x,\cdots ,x_{s}) &=&\int dzJ(x,z)T(\cdots ,z,\cdots ) \\
&=&H_{m}(x)T(\cdots ,x,\cdots )+surface\ terms
\end{eqnarray}
Clearly, $H_{m}=e^{imx}$ acts through multiplication alone. Once again we
see the relationship between $\mathcal{J}$ and $\mathcal{L}_{N},$ since $%
L_{n}=H_{n}D.$ Quite generally, for any weight-$n$ operator $M_{(n)},$ the
product $H_{m}M_{n}$ is a corresponding operator with weight $n+m.$

Again, we define a ``left-moving'' set of operators, $K_{(k)},$ which
anticommute with $D$
\begin{equation}
\{D,\,K_{m}\}=mK_{m}
\end{equation}
but this time have vanishing t'Hooft commutator
\begin{equation}
\lbrack K_{m},K_{n}]_{tH}\equiv K_{m}K_{n}-e^{2i(m-n)x}K_{n}K_{m}=0
\end{equation}
$K_{m}$ therefore takes the form
\begin{equation}
K_{n}=e^{inx}\delta (x+y)
\end{equation}
in the continuous representation. The corresponding integrated operator, $%
I_{m}$ inverts the sign of the tensor field it acts on, so it must be
written as the product of a phase and a parity operator. Let
\begin{equation}
(PT)_{r}(x_{r})=T(-x_{r})
\end{equation}
for the $r^{th}$ slot of any rank tensor $T.$ Then $I_{m}=e^{imx}P.$

The full collection of operators $M_{(k)},N_{(k)},J_{(k)}$ and $K_{(k)}$
together form a t'Hooft-Virasoro quantum algebra. The full algebra is given
in Appendix 2.

In subsequent sections we will employ the mixed representation,
\begin{equation}
M=\sum \alpha _{s}^{k}e^{is(x-y)}
\end{equation}
in addition to the matrix and continuous representations used so far. This
allows greater generality. Often it is convenient to let the mixed and
continuous representations differ by a factor of $2\pi ,$ for example by
writing $D$ as $-i\partial _{x}\delta (x-y)$ in the continuous
representation but as $\sum se^{is(x-y)}$ in the mixed representation. The
extra constant factor generally causes no difficulty.

To conclude this section, we note how neatly the complex, continuous
representation reflects the weight tower structure. Phase transformations
act to change the weights of fields, while a differential representation of
the Virasoro algebra emerges in a natural way from the noncommuting
transformations. Further details on the relationship between units, length
scales and the continuous representation by phases are given in Appendix 3.

We now turn to an examination the tangent space, including a look at the
equivalence relations which define the tangent tower, and a study of the
central charges which arise as a result of the projective character of each
of the ``floors'' of the weight tower.

\section{Projective structure}

\smallskip While the full algebra of transformations $\mathcal{M}$ acts on
the weighted tangent spaces of a scale-invariant geometry, not all of these
transformations are physically meaningful because parallel vectors are
equivalent up to a change of gauge. Since physical laws must be gauge
invariant, those laws must be unchanged if they are rewritten in terms of
any different representative of a given projective equivalence class. This
is why the weight tower is defined using projective Minkowski spaces as the
floors of the tower. In fact, the whole projective tower structure defines
an equivalence relation: any two Lorentz-parallel vectors of the same weight
are equivalent. In this section, we turn this last statement around. We will
look at a heirarchy of three equivalence relations, and the implications
each has for tangent structure, operator equivalence, and Lie algebras of
operators.

To define these three equivalence relations, we must first characterize the
full tangent space, $V$, in a way unbiased regarding the presence of a tower
structure. We therefore assume that $V$ has a countable, definite weight
basis and that each definite weight subspace is $d$-dimensional. While
weaker assumptions are possible, this one guarantees that we know how to
evaluate $D$ on any vector, and is consistent with the physical models we
are trying to describe. Given the definite weight basis $\{w_{m},e_{m\mu
}^{\quad a}\}$, we can construct definite weight projection operators,
either as the outer product of basis vectors such as
\begin{equation}
P_{m}=w_{m}\otimes w_{m}
\end{equation}
or more formally using $D$,
\begin{equation}
P_{m}=\prod_{k\neq m}\frac{D-k}{m-k}
\end{equation}
Given such a basis and a complete set of projection operators, we can
examine equivalence relations on $V.$

The hierarchy of equivalence relations arises from different ways of
specifying when two vectors are parallel. The weakest relation defines two
vectors as equivalent if they are Lorentz-parallel. Thus, two vectors $v_{1}$
and $v_{2}$ in the tangent space at a point $P$ are \textit{Weyl equivalent,
}$v_{1}\cong _{Weyl}v_{2}$, if for every integer $m$ and in each gauge there
exists a number $\lambda _{m,\varphi }\neq 0$  such that
\begin{equation}
P_{m}v_{1}=\lambda _{m,\varphi }P_{m}v_{2}
\end{equation}
The numbers $\lambda _{m,\varphi }$ may be both gauge- and $m$-dependent.
Weyl equivalence recognizes only the different Lorentz directions with no
regard for weight, so the $V$ modulo Weyl equivalence is simply projective $d
$-dim Minkowski space. Weyl equivalence makes no distinction between
parallel vectors of different scaling weights. As a result, the tower
structure is not present.

Strengthening Weyl equivalence slightly, we define $v_{1}$ and $v_{2}$ to be
\textit{weakly equivalent,}
\begin{equation}
v_{1}\cong _{W}v_{2}
\end{equation}
if for every integer $m$ there exists a nonzero number $\lambda _{m}$ such
that
\begin{equation}
P_{m}v_{1}=\lambda _{m}P_{m}v_{2}
\end{equation}
where $\lambda _{m}$ may depend on the weight $m,$ but must be independent
of the gauge. This will be the case if there exists a nondegenerate
gauge-invariant $(0,2)$ tensor $\Lambda $ such that
\begin{equation}
v_{1}=\Lambda v_{2}  \label{weak equivalence}
\end{equation}
$\Lambda $ can be gauge-invariant only if it is of weight zero. Therefore,
by eq.(\ref{generic weighted op}) it must have components
\begin{equation}
(\Lambda )_{\quad m}^{n}=\lambda _{n}\delta _{\quad m}^{n}
\end{equation}
as may also be seen directly from the invariance of eq.(\ref{weak
equivalence}). The eigenvalues, $\lambda _{n}$ may be arbitrary nonzero
numbers, possibly different for each weight in the tower, so that the set of
allowed $\Lambda $s forms a rather large group, $(R-\{0\})^{J}.$
Nonetheless, two definite-weight vectors of different weights are never
weakly equivalent, so weak equivalence is sufficient to define a tower
structure. When restricted to vectors of definite weight, weak equivalence
clearly implies Weyl equivalence, though Weyl equivalence is not defined for
more general tower vectors.

The tower structure arising from weak equivalence is extremely simple.
Computing the quotient, $V$ modulo weak equivalence, we find that two $(0,1)$
tensors are equivalent if and only if they have the same \textit{distribution%
} of nonvanishing components. Every equivalence class of such weight vectors
contains one representative with all of its components equal to either zero
or one. For this reason, while weak equivalence is sufficiently strong to
recognize the tower structure, it will turn out to strongly limit possible
algebras of weight maps.

Finally, two vectors $v_{1},v_{2,}$ are \textit{strongly equivalent,} $%
v_{1}\cong v_{2}$ if for every integer $m$ there exists a nonzero number $%
\lambda $ such that
\begin{equation}
P_{m}v_{1}=\lambda P_{m}v_{2}
\end{equation}
where $\lambda $ is independent of both the weight and the gauge. Stated
more simply, $v_{1}$ and $v_{2}$ must be parallel, so there exists a gauge
constant $\lambda \neq 0$ such that $v_{1}=\lambda v_{2}$. The
gauge-invariant $\lambda $ is just the measurable ratio of the magnitudes of
$v_{1}$ and $v_{2}.$ Notice that two vectors of different, definite weight
can never be strongly equivalent, so strong equivalence, like weak
equivalence, defines a tower structure. Correspondingly, the ratio of
magnitudes of two vectors of different definite weight does not give a
measurable ratio. In fact, strong equivalence defines the tower structure of
the preceeding sections. The definition requires the vectors to be parallel
both in spacetime direction and in their tower direction. Therefore, if the
components of $v_{1}$ and $v_{2}$ are $(v_{1})_{m}^{\mu }$ and $%
(v_{2})_{m}^{\mu },$ respectively, then $(v_{1})_{m}^{\mu }=\lambda
(v_{2})_{m}^{\mu }$ for all $\mu $ and all $m.$ Note that indefinite weight
vectors \textit{can} be strongly equivalent, and that the subspace of
vectors of definite weight, modulo strong equivalence, forms a projective
Minkowski space.

Strong equivalence implies weak equivalence, since whenever $v_{1}=\lambda
v_{2},$ we can choose $\Lambda $ to be the diagonal matrix with all $\lambda
_{n}$ equal to $\lambda $ so that $\Lambda =\lambda 1$. Conversely, two
vectors which are weakly equivalent and such that $\Lambda =\lambda 1$ for
some $\lambda ,$ are strongly equivalent.

Next, we consider the effect of these equivalence relations on weight maps.
Since Weyl equivalence does not give a tower structure, it also allows no
weight maps and has no algebraic structure beyond the Weyl group. Therefore,
for the remainder of the section, we restrict our attention to strong and
weak equivalence.

While both the weak and strong equivalence classes of equal weight parallel
vectors preserve the tower structure, they lead to very different
equivalence classes. Nonetheless, for definite-weight vectors, there is no
difference between the two - any two vectors which are Lorentz parallel and
of the same weight are both strongly and weakly equivalent. The difference
shows up only for weight superpositions. Consider two superpostions
containing the same distribution of floors of the weight tower such that the
projection to any particular floor gives Lorentz-parallel vectors. Then the
two superpositions will in general be weakly, but not strongly, equivalent.

For example, consider two Lorentz vectors $u_{(n)}$ and $v_{(m)}$ of weights
$n$ and $m,$ respectively. Then the superpositions $au_{(n)}+bv_{(m)}$ and $%
cu_{(n)}+dv_{(m)}$ are weakly equivalent for any nonvanishing $a,b,c$ and $d$
, but are strongly equivalent only if the $2$-vectors $(a,b)$ and $(c,d)$
are parallel. In the weak case, the $0$-weight weight map
\begin{equation}
\Lambda =\left(
\begin{array}{cccccc}
\ddots &  &  &  &  &  \\
& 1 &  &  &  &  \\
&  & \frac{a}{c} &  &  &  \\
&  &  & \frac{b}{d} &  &  \\
&  &  &  & 1 &  \\
&  &  &  &  & \ddots
\end{array}
\right)
\end{equation}
accomplishes the equivalence.

The differences between weak and strong equivalences are even more
pronounced when we consider transformations. We define two weight-$n$
transformations $M_{(n)}$ and $N_{(n)}$ to be strongly or weakly equivalent,
$M_{(n)}\cong N_{(n)}$ or $M_{(n)}\cong_{W} N_{(n)}$ if the vectors $M_{(n)}v
$ and $N_{(n)}v^{\prime }$ are strongly or weakly equivalent whenever $v$
and $v^{\prime }$ are correspondingly equivalent. It is trivial to show that
strongly equivalent maps must be proportional, $M=\lambda N,$ but weakly
equivalent maps are highly degenerate.

The definition of weakly equivalent maps requires a nondegenerate
zero-weight map $\Lambda $ satisfying
\begin{equation}
M_{(n)}v=\Lambda N_{(n)}v^{\prime }=\Lambda N_{(n)}\Lambda _{1}v
\end{equation}
for all $v$ and all $\Lambda _{1},$ or simply
\begin{equation}
M_{(n)}=\Lambda N_{(n)}\Lambda _{1}
\end{equation}
In components this implies the existence of numbers $\lambda _{k},\lambda
_{1k}$ such that
\begin{equation}
(M_{(n)})_{\quad m}^{k}=(\Lambda )_{\quad l}^{k}(N_{(n)})_{\quad
i}^{l}(\Lambda _{1})_{\quad m}^{i}=\lambda _{k}\lambda _{1m}(N_{(n)})_{\quad
m}^{k}
\end{equation}
Since $M_{(n)}$ and $N_{(n)}$ are weight-$n$ maps,
\begin{equation}
(M_{(n)})_{\quad m}^{k}=\sum_{s=-\infty }^{\infty }\alpha _{s}^{(n)}\delta
_{s+n}^{k}\delta _{m}^{s}=\lambda _{k}\lambda _{1m}(N_{(n)})_{\quad
m}^{k}=\sum_{s=-\infty }^{\infty }\beta _{s}^{(n)}\lambda _{k}\lambda
_{1m}\delta _{s+n}^{k}\delta _{m}^{s}
\end{equation}
resulting in
\begin{equation}
\alpha _{s}^{(n)}=\beta _{s}^{(n)}\lambda _{s+n}\lambda _{1s}
\end{equation}
for all $s.$ This expression provides a different $\lambda _{s+n}$ for each
value of $\beta _{s}^{(n)}\lambda _{1s}.$ Since $n$ is fixed, and the
operators $M^{(n)}$ and $N^{(n)}$ are determined by $\alpha _{s}^{(n)}$ and $%
\beta _{s}^{(n)},$ \textit{every two operators of the same (definite) weight
having the same zeros are weakly equivalent.} Consequently, since the matrix
representation of a definite weight operator is a shifted-diagonal matrix,
there is only one operator with a given shift and a given set of zeros, up
to weak equivalence.

The difference between strong and weak equivalence is especially important
when we consider Lie algebras of equivalence classes of operators. For the
moment, let $\cong $ denote either equivalence relation and set $\mathcal{S}
\equiv \{M^{\alpha }\mid \,\alpha \in A\}$ where each $M^{\alpha }$
represents an equivalence class $M^{\alpha }=\{M_{\beta }^{\alpha }\}$ of
operators modulo $\cong $, with $\beta $ indexing the members of each class.
Then $\mathcal{S}$ will be a Lie algebra consistent with $\cong $ provided
\begin{equation}
\lbrack M^{\alpha },M^{\alpha ^{\prime }}]\cong c_{\alpha \alpha ^{\prime
}\alpha ^{\prime \prime }}M^{\alpha ^{\prime \prime }}
\end{equation}
that is,
\begin{equation}
\lbrack M_{\beta _{1}}^{\alpha },M_{\beta _{2}}^{\alpha ^{\prime }}]\cong
c_{\alpha \alpha ^{\prime }\alpha ^{\prime \prime }}M_{\beta _{3}}^{\alpha
^{\prime \prime }}
\end{equation}
for any choices of the representatives $\beta _{1},\beta _{2}$ and $\beta
_{3}.$ If we pick initial representatives $M_{0}^{\alpha },M_{0}^{\alpha
^{\prime }}$ and $M_{0}^{\alpha ^{\prime \prime }}$ such that
\begin{equation}
\lbrack M_{0}^{\alpha },M_{0}^{\alpha ^{\prime }}]=c_{\alpha \alpha ^{\prime
}\alpha ^{\prime \prime }}M_{0}^{\alpha ^{\prime \prime }}
\end{equation}
then there must exist some $\Lambda ^{\prime }$ in the equivalence group
such that
\begin{equation}
\lbrack \Lambda _{1}M_{0}^{\alpha },\Lambda _{2}M_{0}^{\alpha ^{\prime
}}]=c_{\alpha \alpha ^{\prime }\alpha ^{\prime \prime }}\Lambda ^{\prime
}\Lambda _{3}M_{0}^{\alpha ^{\prime \prime }}
\end{equation}
for all $\Lambda _{1},\Lambda _{2}$ and $\Lambda _{3}$. Setting $\Lambda
^{\prime }=\Lambda (\Lambda _{3})^{-1},$ we require existence of $\Lambda $
such that
\begin{equation}
\lbrack \Lambda _{1}M_{0}^{\alpha },\Lambda _{2}M_{0}^{\alpha ^{\prime
}}\rbrack=c_{\alpha \alpha ^{\prime }\alpha ^{\prime \prime }}\Lambda
M_{0}^{\alpha ^{\prime \prime }}  \label{Equivalence commutator}
\end{equation}
for each choice of $\Lambda _{1}$ and $\Lambda _{2}$. For strong
equivalence, this is obviously possible because each $\Lambda =\lambda 1$
commutes with any choices of the operators $M_{0}^{\alpha }$ and we simply
set $\lambda =\lambda _{1}\lambda _{2}.$ In this case, we have the algebras
described in Sec.(2), including $\mathcal{L}_{N}$ and $\mathcal{J}.$
However, for weak equivalence eq.(\ref{Equivalence commutator}) is highly
constraining.

A sufficient condition\footnote{%
The potentially weaker assumption that $\exists \Lambda _{2}^{\alpha }$ such
that $M^{\alpha }\Lambda _{2}=\Lambda _{2}^{\alpha }M^{\alpha }$ can also be
shown to be inconsistent with an algebra of definite weight operators $%
M^{(k)}$ with $k\neq 0.$} for a Lie algebra of operators to satisfy eq.(\ref
{Equivalence commutator}) for weak equivalence is to demand that the $%
\Lambda _{1}$ and $\Lambda _{2}$ operators factor out to the left in the
same way as for strong equivalence, but this requires
\begin{equation}
\lbrack M_{0}^{\alpha },\Lambda _{2}]=0
\end{equation}
The only set of operators which commute with all zero-weight matrices $%
\Lambda _{2}$ are the zero-weight matrices themselves. We therefore consider
the physical implications of the zero-weight algebra.

The algebra (or group) of diagonal matrices is composed of commuting
observables. We may therefore conjecture that they correspond to classical
variables, and that classical physics rests ultimately on the assumption of
weak equivalence. Physically, weak equivalence corresponds to ignoring any
physical information contained in the multiplicative constants of weight
superpositions. It only matters \textit{which} weights are superimposed, not
\textit{how much} of each is present.

By contrast, strong equivalence gives meaning to the relative constants
between definite weight components, just as mixed quantum states depend on
the relative phases between the pure states of which they are composed. The
operator algebra projectively preserved by the strong equivalence relation
is not directly measurable, since not all operators are of zero weight.
However, it is possible to measure combinations of these operators or their
effects, for example by using the Hilbert norm discussed above.

If this interpretation of weak and strong equivalences as determining
classical and quantum operators is correct, then the difference between
classical and quantum physics lies in whether the ultimate dynamical laws
apply to each floor of the tower independently, or to arbitrary-weight
vector superpositions as a whole. If each conformal weight evolves
independently of the others, then the operator structure arising from weak
equivalence will be enough to account for the evolution, so the motion will
be described by classical (i.e., commuting) operators. But if tower
superpositions evolve in a way that reflects relationships between different
conformal weight objects, then strong equivalence and the resulting
non-commuting operators must play a role, and the system will be a more
complicated one. Strong conformal dynamics is described in Sec.(7), where we
show that strongly equivalent parallel transport obeys the Schr\"{o}dinger
equation.

\bigskip

Next we consider an extremely important consequence of projective
equivalence - central charges. In string theory, it is equivalence under
unitary projection that gives the Virasoro algebra its now famous central
charges. The unitary equivalence occurs because the theory has been
quantized. However, \textit{scaling geometries already provide real
projective equivalence without quantization.} We now move to a demonstration
that the same central charges occur. For the remainder of our discussion, we
will assume the tower structure to be determined by strong equivalence.

Following [4], suppose we have a Lie group $G$ which acts on a space, with
elements $a,b,...\in G$. Projective equivalence of physical states on that
space (whether scale or unitary) allows us to use a projective
representation $H$ of $G,$ with $A,B,...\in H,$ such that
\begin{equation}
A(a)B(b)=\lambda C(ab)  \label{Projective rep of group}
\end{equation}
where $\lambda $ is a scale or phase factor. This is a necessary condition
for the Lie algebras of $G$ and $H$ to differ by the addition of central
charges to the commutation relations in $H,$
\begin{equation}
\lbrack H_{a},H_{b}]=c_{ab}^{d}H_{d}+c_{ab}1  \label{central charges}
\end{equation}
while sufficiency depends on whether the central term can be removed by
redefining the generators $H_{a}.$ In eq.(\ref{central charges}) the central
charges $c_{ab}=-c_{ba}$ must satisfy the Jacobi-like identity [4]
\begin{equation}
c_{bc}^{a}c_{ad}+c_{cd}^{a}c_{ab}+c_{db}^{a}c_{ac}=0
\label{charge constraint}
\end{equation}
These relations hold whether the factor $\lambda $ in eq.(\ref{Projective
rep of group}) is real or complex.

For the Virasoro algebra, $c_{mn}^{k}=(m-n)\delta _{m+n}^{k},$ so the
central charges must satisfy
\begin{equation}
(m-n)c_{m+n,k}+(k-m)c_{m+k,n}+(n-k)c_{k+n,m}=0  \label{cmn constraint}
\end{equation}
Now, eq.(\ref{central charges}) is invariant under the replacement
\begin{equation}
\hat{c}_{mn}=c_{mn}+(m-n)\phi _{m+n}
\end{equation}
for any $\phi _{k}.$ This means that it is possible to remove some of the
potential central charges by a shift in the generators. In the present case,
the commutator of $M_{(k)}$ with $D,$
\begin{equation}
\lbrack M_{(k)},D]=-kM_{(k)}+c_{k0}1
\end{equation}
becomes
\begin{equation}
\lbrack \hat{M}_{(k)},D]=-k\hat{M}_{(k)}+(c_{k0}+k\phi _{k})1
\end{equation}
under the replacement $\hat{M}_{(k)}=M_{(k)}+\phi _{k}1.$ Therefore, by
choosing $\phi _{k}=-k^{-1}c_{k0}$ we can eliminate $\hat{c}_{k0}.$ Then
setting $k=0$ in eq.(\ref{cmn constraint}) we find
\begin{equation}
(m+n)c_{mn}=0
\end{equation}
so that the central charge $c_{mn}$ vanishes unless $n=-m.$ To find the
magnitude of the charge (as in [1], p80) we return to eq.(\ref{cmn
constraint} ) with $c_{mn}=A(m)\delta _{m+n}^{0}$%
\begin{equation}
\delta _{k+m+n}\left[ (m-k)A(n)+(n-m)A(k)+(k-n)A(m)\right] =0
\end{equation}
Noting that $A(-m)=-A(m),$ we set $k=1,$ $n=-(m+1)$ to find the recursion
relation
\begin{equation}
A(m+1)=\frac{1}{(m-1)}\left[ (m+2)A(m)-(2m+1)A(1)\right]
\end{equation}
The solution,
\begin{equation}
A(m)=am^{3}+bm
\end{equation}
dependent on two initial values, $A(1)\equiv a+b$ and $A(2)\equiv 8a+2b,$ is
easily checked by induction.

Finally, we note that a projective representation also allows the
introduction of central charges into the commuting Lie algebra $\mathcal{J}$%
. The consistency of \textit{abitrary} charges with eq.(\ref{charge
constraint}) is trivial since all of the structure constants vanish. We can
therefore write
\begin{equation}
\lbrack J_{m},J_{n}]=b_{mn}1
\end{equation}
for any constants $b_{mn}=-b_{nm}.$ However, since the right-hand side has
vanishing conformal weight, we must have
\begin{equation}
b_{mn}=b_{m}\delta _{m+n}^{0}
\end{equation}
with $b_{m}=-b_{-m}.$ This remaining constant may be absorbed into the
definition of the $J_{m},$ except for the antisymmetry. The antisymmetry can
be accommodated by scaling $J_{m}$ so that $b_{mn}=m\delta _{m+n}^{0},$
which gives the usual mode algebra
\begin{equation}
\lbrack J_{m},J_{n}]=m\delta _{m+n}^{0}1
\end{equation}
and permits the usual representation of the Virasoro algebra in terms of $%
J_{m}:$
\begin{equation}
L_{n}=\frac{1}{2}\sum\limits_{n=-\infty }^{\infty }J_{m-n}J_{n}
\end{equation}
so the unique fully non-commuting projective Lie algebra is a sub-algebra of
the free algebra of the unique fully commuting projective Lie algebra with
central extension. This representation of $\mathcal{L}_{N}$ in terms of $%
\mathcal{J}$ introduces the usual operator ordering ambiguity for $L_{0}.$

It is intriguing to note that the algebra $\mathcal{J}$ with central charges
has the form of a canonical quantum commutator, suggesting canonical quantum
commutators can be viewed as projective central extensions of commuting
classical operators. In fact, this result is shown in [2] for the case of
canonical commutators for position and momentum variables in the quantum
mechanical limit of the tower structure. The demonstration is comparable for
any canonically (conformally!) conjugate pair of classical variables. Also,
the parallel transport equation for $(0,1)$ tensors on a tower geometry is
equivalent to the Schr\"{o}dinger equation (see [2], and below). \textit{It
appears that the entire canonical quantization procedure can be formulated
in the following way: extend the weak equivalence algebra of classical,
commuting operators by strengthening the equivalence relation and
considering the projective representations allowed by scale invariance, then
let the resulting fields evolve by parallel transport determined by the
connection on the full tangent tower.}

\section{Conformal dynamics}

\smallskip Before considering vector valued weight operators, i.e., rank $%
(1,2)$ tensors, we explore some properties of dynamics on the weight tower.
In order to arrive at stringlike structures, we will need to implement the
steps outlined at the end of Sec.(6) for the description of $(1,2)$ tensor
fields on scale-invariant spacetimes. This procedure will be easier if we
first consider the dynamics of a simple classical variable, the position of
a point particle.

To begin, we need a description by classical commuting variables on the
weakly equivalent tower. There is a step implicit in this, because normally,
classical fields are described using Weyl equivalence rather than weak
equivalence. To see this point clearly, consider the free particle position,
$x^{\mu }.$ Even if we regard $x^{\mu }$ as a weight-1 tower scalar, it is
not an element of the Hilbert space of the weight tower because it is not
self-dual, hence not of positive norm. The simplest way to make it self-dual
is to form a new vector by taking the self-dual part,
\begin{equation}
\eta ^{\mu }(x)=\frac{1}{2}(x_{+1}^{\mu }e^{ix}+x_{-1}^{\mu }e^{-ix})
\end{equation}
Here $x_{+1}^{\mu }$ is the original coordinate, while $x_{-1}^{\mu }$ is
its weight-$(-1)$ dual, and we have used the continuous representation.
Since weight $(-1)$ means the extra part of the field has units of inverse
length, it is trivial to guess that the quantity $x_{-1}^{\mu }$ should be
identified with a multiple of $\frac{p^{\mu }}{\hbar }$. The proportionality
constant turns out to require a factor of $i$ because of the difference
between the zero signature of the Killing metric of the conformal group and
the $+4$ signature of phase space, but we need not go into this here. A more
detailed discussion is given in [2].

The essential point here is that the restriction to self-dual fields
required by the conformal structure automatically introduces a combination
of position and momentum as the principal variable of the theory. This gives
us a deeper understanding of the origin of the uncertainty principle. We now
see the coupled evolution of position and momentum as arising from our need
to use an inner product to form conformally invariant scalars as candidates
for measurement, and from the convergence criterion for vectors needed even
to define the weight tower.

We can easily regard $\eta ^{\mu }$ as a vector on either a weakly
equivalent or a strongly equivalent weight tower. Choosing strong
equivalence as the most predictive (and also the experimentally verified)
tower structure, the final step is to look at parallel transport of $\eta
^{\mu }$ on the spacetime manifold, using the tower connection. As shown in
[2], parallel transport along a curve with tangent $u^{\mu }$ is equivalent
to solving the Lorentz-covariant Schr\"{o}dinger equation. To see how this
arises, consider the form of the covariant exterior derivative of a tower
vector. For any gauge covariant derivative,
\begin{equation}
\mathbf{D}\phi =\mathbf{d}\phi +\phi \mathbf{A}
\end{equation}
where the gauge vector $\mathbf{A}$ is a Lie algebra valued $1$-form. For
dilations, $\mathbf{A}$ is given by the Weyl vector
\begin{equation}
\mathbf{\theta }=\mathbf{d}x^{\mu }\theta _{\mu }D
\end{equation}
where $D$ is the dilation operator. Using our matrix expression
\begin{equation}
(D)_{\quad n}^{m}=\sum_{s=-\infty }^{\infty }s\delta _{s}^{m}\delta _{n}^{s}
\end{equation}
we see that when $\mathbf{D}$ acts on a weight-$n$ vector $\phi _{(n)}$ we
get the usual formula for the Weyl-covariant derivative given by eq.(\ref
{Weyl cov. derivative}). The tangent tower formalism combines into a single
expression the multiplicity of different Weyl-covariant derivatives
represented by eq.(\ref{Weyl cov. derivative}). Additionally, we can now
differentiate indefinite-weight objects such as $\eta ^{\mu }$:
\begin{equation}
D_{\nu }\eta ^{\mu }=\partial _{\nu }\eta ^{\mu }-\eta ^{\beta }\Gamma
_{\beta \nu }^{\mu }+\eta ^{\mu }\theta _{\nu }D\equiv \mathcal{D}_{\nu
}\eta ^{\mu }+\theta _{\nu }\eta ^{\mu }D
\end{equation}
where $\mathcal{D}_{\nu }$ is the usual covariant derivative using the
Christoffel connection, $\Gamma _{\beta \nu }^{\mu }$. For parallel
transport of $\eta ^{\mu }$ we write
\begin{equation}
u^{\nu }\mathcal{D}_{\nu }\eta ^{\mu }=i\mathcal{H}\eta ^{\mu }
\label{Schrodinger eq}
\end{equation}
where $i\mathcal{H}=$ $u^{\nu }\theta _{\nu }D$ is the component along $%
u^{\nu }$ of the operator-valued gauge vector and we have used $\eta ^{\mu
}D=-D\eta ^{\mu }$ (see eqs.(\ref{DT},\ref{TD})). As mentioned above, the
factor of $i$ arises in changing from the zero signature of the Killing
metric of the conformal group to the signature-$4$ metric of the usual phase
space variables [2]. The identification of $\mathcal{H}$ as the Hamiltonian
operator is consistent with the interpretation given in [2] and [3] and also
in earlier related, but distinct, work [5]. With this identification, eq.(%
\ref{Schrodinger eq}) is the Schr\"{o}dinger equation.

A useful relationship between spacetime and internal parameters is provided
by using the continuous representation of the dilation operator in the
Schr\"{o}dinger equation. Consider the action of $D(x,y)$ on a general
vector, $f(y):$%
\begin{equation}
Df(x)=\int (-i\partial _{x}\delta (x-y))f(y)=-i\partial _{x}f(x)
\end{equation}
or simply
\begin{equation}
D=-i\frac{\partial }{\partial x}  \label{D=idx}
\end{equation}
Combining eq.(\ref{D=idx}) with eq.(\ref{Schrodinger eq}), in flat space so
that $\frac{D}{d\tau }=\frac{d}{d\tau },$ we have
\begin{equation}
u^{\nu }\mathcal{D}_{\nu }\eta ^{\mu }=\frac{d\eta ^{\mu }(\tau ;x)}{d\tau }%
=i(u^{\nu }\theta _{\nu })\frac{\partial }{\partial x}\eta ^{\mu }(\tau ;x)
\label{Map x to tau}
\end{equation}
Since $\theta \equiv u^{\nu }\theta _{\nu }$ is implicitly a function of $%
\tau ,$ eq.(\ref{Map x to tau}) provides an explicit map, $\Phi ,$ unique up
to reparameterization, between the parameter, $x,$ of the continuous
representation, and the proper time coordinate, $\tau ,$ along a curve in
spacetime. This map provides an embedding of the parameter space $[-\pi ,\pi %
]$ into the spacetime manifold, $\mathcal{M},$ which is necessarily $1-1.$
As we shall see in the next section, this same embedding also relates $(1,2)$
tensors to string.

Canonical commutators for conformally conjugate variables hold if we use the
correspondence developed in Appendices 4 and 5 to turn the vector $\eta
^{\mu }$ into a sum of \textit{definite weight} operators. Taking the $\Phi $
into account, we begin with the components of the $(1,1)$-vector $(\eta
^{0},\eta ^{k}),$%
\begin{eqnarray}
x^{0} &=&t=x=i\sum \frac{(-)^{m}}{m}e^{imx} \\
x^{k} &=&x_{1}^{k}e^{ix}+x_{-1}^{k}e^{-ix}\qquad (k=1,2,3)
\end{eqnarray}
and map to the corresponding definite weight operators
\begin{eqnarray}
T &=&tJ_{0}=i\sum \frac{(-)^{m}}{m}J_{m}(x-y)\equiv \sum T_{m} \\
X^{k} &=&x^{k}J_{0}=\frac{x_{1}^{\mu }}{2}J_{1}+\frac{x_{-1}^{\mu }}{2}J_{-1}
\end{eqnarray}
For the spatial components we can the identify conjugate operators
(``position'' and ``momentum'' operators of weight $(+1)$ and $(-1)$,
respectively),
\begin{eqnarray}
\eta _{1}^{k} &=&\frac{1}{2}x_{1}^{k}e^{ix}\sum_{n}e^{in(x-y)}=\frac{%
x_{1}^{k}}{2}J_{1} \\
\eta _{-1}^{k} &=&\frac{1}{2}x_{-1}^{k}e^{-ix}\sum_{n}e^{in(x-y)}=\frac{%
x_{-1}^{k}}{2}J_{-1}
\end{eqnarray}
These operators commute, except for the central term allowed by using a
projective representation, in which case we have the usual canonical
commutators
\begin{eqnarray}
\lbrack \eta _{+1}^{k},\eta _{+1}^{k^{\prime }}] &=&[\eta _{-1}^{k},\eta
_{-1}^{k^{\prime }}]=0 \\
\lbrack \eta _{+1}^{k},\eta _{-1}^{k^{\prime }}] &=&\frac{1}{4}%
x_{1}^{k}x_{-1}^{k^{\prime }}\delta ^{kk^{\prime }}\mathbf{1}
\end{eqnarray}
For the time coordinate, however, there is no simple single-mode operator.
There is instead an infinity of operators $T_{m}$ satisfying
\begin{equation}
\lbrack T_{m},T_{n}]=\frac{1}{m}\delta _{m+n}^{0}\mathbf{1}
\end{equation}
Because the Schr\"{o}dinger equation gives us a timelike embedding of the
parameter space into spacetime, the time operator does not satisfy the same
simple commutation with energy that holds for the spatial components with
momentum, but instead reproduces the entire $\mathcal{J}$ algebra.

We also note briefly the algebra associated with the self-dual form of $\eta
^{\mu }(x).$ Beginning with the self-dual operator
\begin{eqnarray}
\eta ^{\mu }(x,y) &\equiv &\frac{1}{2}\left( \eta ^{\mu }(x-y)+\eta ^{\mu
}(-(x+y))\right)  \\
&=&\frac{1}{2}\sum_{k=-1,1}x_{k}^{\mu }\cos kx\ e^{-iky} \\
&=&\frac{1}{4}\left( x_{+1}^{\mu }e^{i(x-y)}+x_{-1}^{\mu
}e^{-i(x-y)}+x_{+1}^{\mu }e^{-i(x+y)}+x_{-1}^{\mu }e^{i(x+y)}\right)
\end{eqnarray}
we extract the operators
\begin{eqnarray}
J_{1} &=&e^{i(x-y)}  \nonumber \\
J_{-1} &=&e^{-i(x-y)}  \nonumber \\
K_{1} &=&e^{-i(x+y)}  \nonumber \\
K_{-1} &=&e^{i(x+y)}  \label{SD algebra}
\end{eqnarray}
Since the operators $K_{1}$ and $K_{-1}$ are $0$-weight $K$-type operators
they satisfy normal rather than t'Hooft commutation relations. Because of
this, $J_{1,-1},K_{1,-1}$ form a $4$-dim Lie algebra, which can be shown to
be $gl(2,R)$ or $u(1)\times su(1,1).$

\section{String representations of vector valued weight operators}

Before venturing into the \textit{uniqueness }of the relationship between
moving quantized string and the $(1,2)$ tensors, we note that the \textit{%
existence} of such a relationship has already been established. The crucial
elements of string theory have already been described as elements of
scale-invariant geometry: we have the mode algebra, $\mathcal{J},$ and the
Virasoro algebra, both with central charges, acting on Hilbert space. Since
the $(1,2)$ tensors are vector valued weight maps, those maps will be
elements of either the mode or the Virasoro algebras. We could simply
consider the case when the maps are in the mode algebra, and follow the
standard path for describing string states. In this sense, our goal is
accomplished. The motion of strings in spacetime could also be fixed by
fiat, by setting the Lorentz vector part of the tensor to have the
particular properties of a string world sheet. But, as we now show, these
properties follow automatically from the map $\Phi $ defined in Sec.(7)
together with the requirement of measurabity of states. With this goal in
mind, we turn to a study of $(1,2)$ tensors.

Rank $(1,2)$ tensors have the general form $P_{\quad n}^{\mu m}$ or $P^{\mu
}(x,y).$  There are two reasons (aside from looking like string variables)
that the type $(1,2)$ tensor fields are especially important. First, they
may be used to construct arbitrary, higher order tensorial weight maps of
rank $(r,2)$ simply by taking products of maps. Second, they provide a
correspondence between weight maps and vector fields. The correspondence
permits us to move freely back and forth between tensor fields on spacetime
and fields of operators on Hilbert space. We will be particularly interested
in the case where this map is $1-1.$

Our procedure in this section resembles the Hiesenberg picture of quantum
mechanics, in which it is the operators rather than the states which evolve.
We assume that these operators can act on general states, with the
properties of the operators reflected in the result. Before we study the
evolution of these $(1,2)$ operators, we show how to build convergence into
them.

From this Heisenberg point of view, the most general measurable $(1,2)$
tensor is one which produces an element of the self-dual Hilbert space from
a general vector. We therefore ask for the most general $(1,2)$ tensor field
$P^{\mu }(x,y)$ for which $(P^{\mu }f)(x)$ is self-dual for all vectors $%
f(x).$ Such an operator will be called self-dual. Ideally, $P^{\mu }(x,y)$
would also be required to be of definite weight, but we can easily see from
the matrix representation that a self-dual operator must be symmetric about
the $m=0$ row, top to bottom. Such an object cannot be of definite weight,
but can be of definite weight plus anti-weight, a property which we will call%
\textit{\ semi-definite-weight}. Beginning with the general form for a
semi-definite-weight-$k$ operator $P_{(k)}^{\mu }(x,y)$ and a general vector
$f(x)$,
\begin{eqnarray}
P_{(k)}^{\mu }(x,y) &=&e^{ikx}\sum_{m=-\infty }^{\infty }\left( \beta ^{\mu
m}e^{im(x-y)}+\gamma ^{\mu m}e^{im(x+y)}\right)  \\
f(x) &=&\sum_{m=-\infty }^{\infty }c^{m}e^{imx}
\end{eqnarray}
the product is
\begin{eqnarray}
(P_{(k)}^{\mu }f)(x) &=&\frac{1}{2\pi }e^{ikx}\int_{-\pi }^{\pi }dy\left(
\beta ^{\mu m}e^{im(x-y)}+\gamma ^{\mu m}e^{im(x+y)}\right) c^{k}e^{iky} \\
&=&\sum_{m}e^{ikx}\left( \beta ^{\mu m}c^{m}e^{imx}+\gamma ^{\mu
m}c^{-m}e^{imx}\right)  \\
&=&\sum_{n}\left( \beta ^{\mu ,n-k}c^{n-k}+\gamma ^{\mu ,n-k}c^{-n+k}\right)
e^{inx}
\end{eqnarray}
The product is therefore self-dual provided
\begin{equation}
\left( \beta ^{\mu ,n-k}c^{n-k}+\gamma ^{\mu ,n-k}c^{-n+k}\right) =\left(
\beta ^{\mu ,-n-k}c^{-n-k}+\gamma ^{\mu ,-n-k}c^{n+k}\right)
\end{equation}
for all $c^{n}.$ Now, $c^{n-k},c^{-n+k},c^{-n-k}$ and $c^{n+k}$ are
independent, leaving no solutions, unless either $n=0$ or $k=0.$ In the
first case the expression holds identically, while for $k=0$ we must have
\begin{equation}
\beta ^{\mu n}=\gamma ^{\mu ,-n}\equiv \frac{1}{2}\alpha ^{\mu n}
\end{equation}
for all $n\neq 0.$ The most general self-dual, semi-definite weight operator
is therefore either
\begin{eqnarray}
P_{(0)}^{\mu }(x,y) &=&\alpha ^{\mu 0}+\frac{1}{2}\sum_{m\neq 0}\alpha ^{\mu
m}\left( e^{im(x-y)}+e^{-im(x+y)}\right)  \\
&=&\sum_{m=-\infty }^{\infty }\alpha ^{\mu m}\cos (mx)\ e^{-imy}
\end{eqnarray}
if the semi-definite weight is zero, or
\begin{equation}
F_{(k)}^{\mu }(x,y)=\lambda _{k}^{\mu }e^{iky}+\lambda _{-k}^{\mu
}e^{-iky}=F_{(k)}^{\mu }(y)   \label{F(y)}
\end{equation}
if the operator is of semi-definite weight $k.$ In order to predict
measurable quantities, we will be interested in the zero-semi-weight
operators.

Self-duality is not the only condition required for positive norm, because $%
P_{(0)}^{\mu }(x,y)$ is a vector field as well as an operator. The full norm
therefore involves the Minkowski inner product as well as the Hilbert inner
product. Clearly, requiring $P_{(0)}^{\mu }(x,y)$ to be either spacelike or
timelike will keep the sign definite. Of these, only the (forward) timelike
vectors are closed under addition. More importantly, we want to parallel
transport states in the direction of $P_{(0)}^{\mu }(x,y)$ using eq.(\ref
{Schrodinger eq}), which requires that $P_{(0)}^{\mu }(x,y)$ be timelike.

\smallskip

Next we consider the evolution states along $P_{(0)}^{\mu }(x,y).$ For this
purpose we require the integral curves, $X^{\mu }(\tau ;x,y)$ of $%
P_{(0)}^{\mu }(x,y).$ While integral curves are well-defined for vector
fields, the development of the operator part of $X^{\mu }(\tau ;x,y)$
remains to be specified. To this end we require the particular integral
operator-curve $X^{\mu }(\tau ;x,y)$ such that states $X^{\mu }f$ evolve
along $P^{\mu }$ by parallel transport. We will call such a set of curves
the \textit{integral curves by parallel transport}, or more simply, the
\textit{integral transport curves.}

To be integral curves of $P^{\mu },$ the curves must satisfy\footnote{%
The proof that $X^{\mu }$ has semi-weight zero is simply to write is as an
explicit sum of terms in $(x-y)$ or $(x+y),$%
\begin{equation}
X_{(0)}^{\mu }(x,y)=x^{\mu }+\frac{1}{2}\alpha _{0}^{\mu }\left(
(x+y)-(x-y)\right) +i\sum \frac{1}{m}\alpha _{m}^{\mu }e^{im(x-y)}+i\sum
\frac{1}{m}\alpha _{m}^{\mu }e^{-im(x+y)}
\end{equation}
Each of these terms either commutes or anticommutes with $D.$}
\begin{equation}
\frac{d}{d\tau }X_{(0)}^{\mu }(\tau ;x,y)=P_{(0)}^{\mu }(x,y)
\end{equation}
Then, for arbitrary $f(x)$, we require the parallel transport equation, eq.(%
\ref{Schrodinger eq}) of Sec.(7), to evolve the tower vector $X^{\mu }f$:
\begin{equation}
u^{\nu }\mathcal{D}_{\nu }(X_{(0)}^{\mu }f)+(X_{(0)}^{\mu }f)\theta D=0
\end{equation}
where we have written $\theta =\theta (\tau )$ for $u^{\nu }\theta _{\nu }$.
Commuting $fD=-Df,$ and letting all spacetime dependence reside in the
operator valued coordinate, $X_{(0)}^{\mu }=X_{(0)}^{\mu }(\tau ;x,y)$
rather than in $f=f(x),$ the transport equation becomes
\begin{equation}
\frac{1}{\theta (\tau )}\ \frac{\mathcal{D}X_{(0)}^{\mu }}{d\tau }\
f=X_{(0)}^{\mu }Df
\end{equation}
or since $f$ is arbitrary,
\begin{equation}
\frac{1}{\theta (\tau )}\ \frac{\mathcal{D}X_{(0)}^{\mu }}{d\tau }\
=X_{(0)}^{\mu }D
\end{equation}
Finally, applying eq.(\ref{TD}) and reparameterizing the spacetime path, we
have
\begin{equation}
\frac{\mathcal{D}X_{(0)}^{\mu }(x,y)}{d\lambda }\equiv \frac{1}{\theta (\tau
)}\ \frac{\mathcal{D}X_{(0)}^{\mu }(x,y)}{d\tau }\ =i\frac{\partial }{%
\partial y}X_{(0)}^{\mu }(x,y)
\end{equation}
In flat spacetime, with $\frac{\mathcal{D}}{d\lambda }\rightarrow \frac{d}{%
d\lambda },$ we can identify
\begin{equation}
\frac{d}{d\lambda }=\frac{1}{\theta (\tau )}\frac{d}{d\tau }=i\frac{\partial
}{\partial y}
\end{equation}
or
\begin{equation}
y=i\int \theta (\tau )d\tau =i\lambda   \label{y=lambda}
\end{equation}
that is, we embed the $y$-parameter into $\mathcal{M}$ as before. Succinctly
put, $X^{\mu }(\tau ;x_{0},y)$ is the lifting of the integral curves $X^{\mu
}(\tau )$ of $P^{\mu }(\tau ;x_{0},y)$ into the tower bundle in such a way
that $y=i\lambda $. The factor of $i$ in these expressions is again due to
the use of the physical variable $u^{\mu }=\frac{dx^{\mu }}{d\lambda }$
instead of the geometric variable $iu^{\mu }$ which arises from conformal
gauge theory. In the geometric picture, this is a real-valued embedding [3]
and the factor $i$ is part of the signature of the space.

Now suppose that in some region, $\mathcal{S}$, of flat spacetime (so that $%
\frac{\mathcal{D}}{d\tau }\rightarrow \frac{d}{d\tau })$ we have a
non-constant, Lorentz-timelike, zero-weight $C^{\infty }$ operator field $%
P_{(0)}^{\mu }(x^{\alpha };x,y)$. Fix $x=x_{0},$ let an initial point $%
\mathcal{P}$ have spacetime coordinates $x^{\mu },$ and suppose $%
X_{(0)}^{\mu }(\lambda ;x_{0},y)$ is the integral transport curve of $%
P_{(0)}^{\mu },$ that is,
\begin{equation}
\frac{dX_{(0)}^{\mu }}{d\lambda }=i\frac{\partial X_{(0)}^{\mu }}{\partial y}%
=P_{(0)}^{\mu }  \label{Integral curves}
\end{equation}
Eq.(\ref{Integral curves}) is easily integrated using the explicit form
\begin{equation}
P_{(0)}^{\mu }(x^{\alpha };x,y)=i\sum \alpha ^{\mu m}(x^{\alpha })\cos mx\
e^{-imy}
\end{equation}
for $P_{(0)}^{\mu },$ giving the integral transport curve,
\begin{equation}
X_{(0)}^{\mu }(x_{0},y=i\lambda )=x^{\mu }(\mathcal{P})+\alpha ^{\mu
0}y+i\sum_{m\neq 0}\frac{1}{m}\alpha ^{\mu m}\cos (mx_{0})\ e^{-imy}
\end{equation}
which is precisely eq.(\ref{Open string}) for the open string evaluated
along a timelike curve, $\sigma =x_{0}.$ The integration produces a
self-dual vector field. Notice that the linear function of $y$ (or the
polynomial that arises from further integration) is consistent with
self-duality because any function of $y$ alone is a self-dual operator (see
also eq.(\ref{F(y)})). A function of $x$ alone is not self-dual.

When we perform the integral above for each point $x$ in $\mathcal{S},$ we
get a series of curves $X_{(0)}^{\mu }(x,y)$ with tangent vectors $%
P_{(0)}^{\mu }(x,y).$ This gives an injective map $\Phi ^{\mu }$ defined by
\begin{eqnarray}
\Phi ^{\mu } &:&\mathcal{S}\rightarrow \mathcal{M} \\
\Phi ^{\mu }(x,y) &\equiv &X^{\mu }(x,y)
\end{eqnarray}
The injected set has dimension $\leq 2.$ We now prove that $\Phi ^{\mu }$ is
an immersion of the entire region $\mathcal{N}_{\pi }\equiv (0,\pi )\times
(0,\pi )\subset \mathcal{N}$ into $\mathcal{M}$ (and therefore of dimension $%
2$).

At a general value of the pair $(x,y)$ consider the tangent space to the
injected set $X_{(0)}^{\mu }(x,y).$ This space $\mathcal{V}_{2}$ consists of
all vectors
\begin{eqnarray}
V^{\mu }(\alpha ,\beta ) &=&\left( \alpha \frac{\partial }{\partial x}+\beta
\frac{\partial }{\partial y}\right) X_{(0)}^{\mu }(x,y) \\
&=&\alpha \left( -i\sum_{m\neq 0}\alpha ^{\mu m}\sin (mx)\ e^{-imy}\right)
+\beta \left( \alpha _{0}^{\mu }+\sum_{m\neq 0}\alpha ^{\mu m}\cos (mx)\
e^{-imy}\right)
\end{eqnarray}
which will be $2$-dimensional provided the basis vectors
\begin{eqnarray}
Q^{\mu } &=&\left( -i\sum_{m=-\infty }^{\infty }\alpha ^{\mu m}\sin (mx)\
e^{-imy}\right)   \label{Q} \\
P^{\mu } &=&\sum_{m=-\infty }^{\infty }\alpha ^{\mu m}\cos (mx)\ e^{-imy}
\end{eqnarray}
are independent.

We begin by looking at the inner product $P^{\mu }Q_{\mu }$:
\begin{equation}
P^{\mu }Q_{\mu }=-i\sum_{m,n=-\infty }^{\infty }\alpha ^{\mu m}\alpha \alpha
_{\mu }^{n}\cos (mx)\ \sin (nx)\ e^{-i(m+n)y}  \label{PQ}
\end{equation}
The action of $P^{\mu }Q_{\mu }$ on an arbitrary self-dual vector $f=\sum
c^{k}e^{ikx}$ is given by
\begin{eqnarray}
P^{\mu }Q_{\mu }f &=&-\frac{i}{2\pi }\int dy\sum_{k,m,n=-\infty }^{\infty %
}\alpha ^{\mu m}\alpha _{\mu n}\cos (mx)\ \sin (nx)\ e^{-i(m+n-k)y}c^{k} \\
&=&-i\sum_{m,n=-\infty }^{\infty }\alpha ^{\mu m}\alpha _{\mu
}^{n}c^{m+n}\cos (mx)\ \sin (nx) \\
&=&-i\sum_{m,n=-\infty }^{\infty }\alpha _{m}^{\mu }\alpha _{\mu
n}c^{-m-n}\cos (-mx)\ \sin (-nx) \\
&=&+i\sum_{m,n=-\infty }^{\infty }\alpha ^{\mu m}\alpha _{\mu
}^{n}c^{m+n}\cos (mx)\ \sin (nx) \\
&=&0
\end{eqnarray}
where we use the self-dual representation $\alpha ^{\mu m}=\alpha _{m}^{\mu }
$ and $c^{k}=c_{k}$ in the penultimate step (see Appendix 4). Therefore, $%
P^{\mu }Q_{\mu }=0$ on all measurable vectors, and so is equivalent to the
zero operator. Now, setting eq.(\ref{PQ}) for $P^{\mu }Q_{\mu }$ to zero
explicitly, we can multiply both sides by $e^{iky}$ and integrate over $y$
to find
\begin{eqnarray}
0 &=&-i\sum_{m=-\infty }^{\infty }\alpha ^{\mu m}\alpha _{\mu }^{k-m}\cos
(mx)\ \sin (k-m)x \\
&=&-\frac{i}{2}\sum_{m=-\infty }^{\infty }\alpha ^{\mu m}\alpha _{\mu
}^{k-m}\left( \sin kx-\sin (2m-k)x\right)
\end{eqnarray}
Now, multiplying by $\sin nx$ and integrating over $x$ gives two nonzero
cases. When $n=k,$ we have\footnote{%
Once the $P^{\mu m}$ are represented with central charges, the $\alpha ^{\mu
m}$ acquire true operator status. Then eq.(\ref{Physical states})
corresponds to the physical state conditions $L_{m}\phi =0$ for string.}
\begin{equation}
\sum_{m=-\infty }^{\infty }\alpha ^{\mu m}\alpha _{\mu }^{k-m}=0
\label{Physical states}
\end{equation}
while $n=2m-k$ gives an equivalent result. Working backward, we write
\begin{eqnarray}
0 &=&\sum_{m=-\infty }^{\infty }\alpha ^{\mu m}\alpha _{\mu }^{k-m}\cos kx \\
&=&\sum_{m=-\infty }^{\infty }\alpha ^{\mu m}\alpha _{\mu }^{n}\cos kx\
\delta _{k-m}^{n} \\
&=&\sum_{m=-\infty }^{\infty }\alpha ^{\mu m}\alpha _{\mu }^{n}\cos (m+n)x\
\frac{1}{2\pi }\int e^{-i(n+m)z}e^{ikz}dz
\end{eqnarray}
This expression, multiplied by $e^{-iky},$ then summed on $k,$ gives a $%
\delta $-function, $2\pi \delta (y-z).$ Performing the $z$-integration then
leads to
\begin{equation}
0=\sum_{m=-\infty }^{\infty }\alpha ^{\mu m}\alpha _{\mu }^{n}\cos (m+n)x\
e^{-i(n+m)y}
\end{equation}
This same expression arises if we compute $P^{2}+Q^{2},$%
\begin{eqnarray}
P^{2}+Q^{2} &=&\sum_{m,n=-\infty }^{\infty }\alpha ^{\mu m}\alpha _{\mu
}^{n}(\cos (mx)\ \cos (nx)\ -\sin (mx)\ \sin (nx))\ e^{-i(m+n)y} \\
&=&\sum_{m=-\infty }^{\infty }\alpha ^{\mu m}\alpha _{\mu }^{k-m}\cos
(m+n)x\ e^{-i(n+m)y} \\
&=&0
\end{eqnarray}
so that $P^{\mu }P_{\mu }=-Q^{\mu }Q_{\mu }<0,$ where the inequality holds
because $P^{\mu }$ is our original timelike $(1,2)$ tensor field. Therefore,
$Q^{\mu }$ is a nonvanishing spacelike $(1,2)$ vector field orthogonal to $%
P^{\mu },$ and the immersion is established at every point of $(0,\pi )%
\times (-\pi ,\pi ).$ At the endpoints $x=0$ or $\pi ,$ $Q^{\mu }$ vanishes
because $\sin n\pi $ vanishes$.$ The map $\Phi ^{\mu }$ is therefore an
immersion of the region $\mathcal{N}_{\pi }$ into $\mathcal{M}$, where it is
convenient to restrict $y$ to the same interval as $x,$ setting $\mathcal{N}%
_{\pi }=(0,\pi )\times (0,\pi ).$

As a consequence of the immersion $\Phi ^{\mu },$ the Lorentzian metric
structure of $\mathcal{M}$ induces a Lorentzian structure on $\mathcal{N}_{%
\pi }$. To establish the Lorentzian structure on $\mathcal{N}_{\pi }$, we
start with vectors on $\mathcal{N}_{\pi }$, using the parameters $(x,y)$ as
a basis. Letting
\begin{equation}
\sigma ^{a}\equiv (x,y)
\end{equation}
we have the general vector
\begin{equation}
v=v^{a}\frac{\partial }{\partial \sigma ^{a}}
\end{equation}
The map $\Phi ^{\mu }=X^{\mu }$ $(x,y)$ pushes this vector forward to a
vector $V$ tangent to the immersed $2$-surface in $\mathcal{M}$,
\begin{equation}
V=v^{a}\frac{\partial X^{\mu }}{\partial \sigma ^{a}}\frac{\partial }{%
\partial x^{\mu }}=V^{\mu }\frac{\partial }{\partial x^{\mu }}
\end{equation}
The induced metric on the immersed $2$-surface is now pulled back from $%
\mathcal{M}$ by defining $v^{a}v_{a}$ to be equal to $V^{\mu }V_{\mu }$,
\begin{eqnarray}
V^{\mu }V_{\mu } &=&g_{\mu \nu }V^{\mu }V^{\nu }=g_{\mu \nu }\left( v^{a}%
\frac{\partial X^{\mu }}{\partial \sigma ^{a}}\right) \left( v^{b}\frac{%
\partial X^{\nu }}{\partial \sigma ^{b}}\right)  \\
&\equiv &h_{ab}v^{a}v^{b}
\end{eqnarray}
so $h_{ab}$ is given by
\begin{equation}
h_{ab}=g_{\mu \nu }\frac{\partial X^{\mu }}{\partial \sigma ^{a}}\frac{%
\partial X^{\nu }}{\partial \sigma ^{b}}=\left(
\begin{array}{cc}
P^{2} & P^{\mu }Q_{\mu } \\
P^{\mu }Q_{\mu } & Q^{2}
\end{array}
\right) =Q^{2}\eta _{ab}
\end{equation}
Since the tangent space to $\Phi ^{\mu }(\mathcal{N}_{\pi })$ is just $%
\mathcal{N}_{\pi }$, $h_{ab}$ provides a metric on both the immersed $2$
surface and on $\mathcal{N}_{\pi }$. We can give this expression manifest $2$%
-dim conformal invariance by using the inverse metric, $h^{ab},$ to write
\begin{equation}
2=h^{ab}h_{ab}=h^{ab}\frac{\partial X^{\mu }}{\partial \sigma ^{a}}\frac{%
\partial X_{\mu }}{\partial \sigma ^{b}}
\end{equation}
and therefore
\begin{equation}
h_{ab}=\frac{2\partial _{a}X^{\mu }\partial _{b}X_{\mu }}{h^{cd}\partial %
_{c}X^{\nu }\partial _{d}X_{\nu }}  \label{World sheet metric}
\end{equation}

Returning to $\mathcal{N}_{\pi },$ we next look at the null vectors,
\begin{equation}
h(v,v)=Q^{2}\eta _{ab}v^{a}v^{b}=0
\end{equation}
Clearly, the two null directions are
\begin{equation}
\frac{\partial }{\partial x}\pm\frac{\partial }{\partial y}
\end{equation}
Thus the induced Lorentz metric makes rigorous the ``left-moving'' and
``right-moving'' nomenclature for the $(x+y)$ and $(x-y)$ Fourier modes. In
terms of $h_{ab},$ these null modes are left- and right- moving waves, and $%
P_{(0)}^{\mu }(x,y),\ Q^{\mu }(x,y)$ and $X_{(0)}^{\mu }(x,y)$ all satisfy
the wave equation,
\begin{equation}
\sqcap \!\!\!\!\!\sqcup _{h}X_{(0)}^{\mu }(x,y)\equiv \frac{1}{\sqrt{-h}}%
\partial _{a}\left( \sqrt{-h}h^{ab}\partial _{b}X_{(0)}^{\mu }(x,y)\right) =0
\label{String wave eq.}
\end{equation}
Similarly, it is a simple exercise to check directly that a Lorentz
transformation $\Lambda $ on $\mathcal{M}$ induces a corresponding Lorentz
transformation on $\mathcal{N}_{\pi }$.

The action of $\Phi ^{\mu }$ on $\mathcal{N}_{\pi }$ gives the world sheet
\begin{equation}
X_{(0)}^{\mu }(x,y)=x^{\mu }(\mathcal{P})+\alpha _{0}^{\mu }y+i\sum_{m\neq 0}%
\frac{1}{m}\alpha ^{\mu m}\cos (mx)\ e^{-imy}  \label{String Rep}
\end{equation}

Eq.(\ref{String Rep}) will be called the \textit{string representation} of
the zero-semi-weight self-dual $(1,2)$ tensor. While eq.(\ref{String Rep})
has the appearance of a standard Fourier series, it differs in an important
respect, namely, that the parameters $\sigma ^{a}$ are labels for the space
of operators on the weight tower. The explicit $m$-dependence of the
coefficient $\alpha ^{\mu m}$ determines which algebra the operator $\alpha
^{\mu m}\cos (mx)\ e^{-imy}$ belongs to. By contrast, the amplitude, $\alpha
^{\mu m}$ \textit{is} the entire operator in the usual quantized string
expansion, eq.(\ref{Open string}). However, this picture changes when we
look at the details of the mode algebra.

To see the emergence of the usual mode algebra, i.e., the canonical
commutators of the mode amplitudes, we follow the example of the single
particle above, mapping to a set of definite weight operators. Following the
steps of Appendix 5 we insure a unique mapping by starting with the
self-dual operator $P_{0}^{\mu }(x,y)=v^{\mu }(x-y)+v^{\mu }(x+y)$,
extracting the $(1,1)$ vector $v^{\mu }(x)=\sum \alpha ^{\mu m}e^{imx},$
then mapping from $v^{\mu }(x)$ to the associated definite weight series,
\begin{equation}
V^{\mu }\equiv v^{\mu }(x)J_{0}(x,y)=\sum_{m}\alpha ^{\mu m}\
J_{m}=\sum_{m}V_{(m)}^{\mu }\
\end{equation}
These operators therefore commute except for possible central extensions.
Using the self-dual form with the maximal central extension as developed in
Appendix 5, the commutation algebra is the usual string mode algebra,
\begin{equation}
\lbrack V_{(m)}^{\mu },V_{(n)}^{\nu }]=m(V\cdot V)\delta _{m+n}^{0}\eta
^{\mu \nu }\mathbf{1}
\end{equation}
provided only that we normalize the amplitudes $(V\cdot V)$ to unity.

At this point, the question arises of how to represent the mode algebra with
central charges. The Fourier exponentials, while operators, all commute and
therefore may be treated as regular numbers. Consequently, there is no harm
in treating the $\alpha ^{\mu m}$ as the projective representation of the
algebra, since a different representation is required anyway. There remains
the difference from standard string theory, that here the mode algebra, like
any set of canonical commutators, is seen only in the associated
definite-weight algebra. Nonetheless, the associated algebra may be used in
the standard way to specify physical states, construct the Virasoro and
Poincar\'{e} generators and so on. Moreover, as we summarize in our final
theorem below, both expressions describe extremal world sheets moving in
spacetime.

Before moving to the theorem, we note that the operators present in eq.(\ref
{String Rep}) for the string representation of $X^{\mu }(x,y)$ also form the
self-dual algebra associated with the mode algebra, in a way perfectly
parallel to the $gl(2,R)$ algebra associated with the self-dual commutators
of Sec.(6), eqs.(\ref{SD algebra}). This time the algebra, with basis
\begin{equation}
e^{im(x-y)},e^{im(x+y)}
\end{equation}
obviously generates the analytic functions on $\mathcal{N}$. For each pair $%
(m,-m)$ the algebra contains a copy of $gl(2,R).$ The restriction of this
algebra to the combination of generators which actually occurs in $X^{\mu },$
namely
\begin{equation}
S_{m}=\cos mx\ e^{-imy}
\end{equation}
generates those functions on $\mathcal{N}$ which have cosine expansions for
the $x$ variable.

We conclude by showing that the $(1,2)$ tensor field $X^{\mu }$ satisfies
the $X^{\mu },$ $h_{ab},$ and endpoint variations of the string action
\begin{equation}
S=\int \sqrt{h}h^{ab}g_{\mu \nu }\partial _{a}\alpha ^{\mu }\partial
_{b}\alpha ^{\nu }
\end{equation}
This means that like string theory, the conformal dynamics of $(1,2)$
tensors describes an extremal world sheet moving in spacetime.

\smallskip

\smallskip

\begin{theorem}
4: The integral transport curves of any timelike, zero-weight, $C^{\infty },$
self-dual operator field $P_{(0)}^{\mu }(x^{\alpha };x,y)$ provide an
immersion of $\mathcal{N}_{\pi }$ into $\mathcal{M}$, causal in the
pull-back of the spacetime metric, such that the immersed $2$-surface
extremizes the action
\begin{equation}
S=\int \sqrt{h}h^{ab}g_{\mu \nu }\partial _{a}X^{\mu }\partial _{b}X^{\nu }
\end{equation}
with respect to $\delta h^{ab}$ and $\delta X^{\mu },$ including the
boundary condition.
\end{theorem}

\smallskip

\begin{description}
\item[Proof]  The variation $\delta X^{\mu }$ leads to
\begin{equation}
\sqcap \!\!\!\!\!\sqcup _{h}X_{(0)}^{\mu }(x,y)=0,
\end{equation}
where the surface term vanishes provided
\begin{equation}
Q^{\mu }(0)=Q^{\mu }(\pi )=0,
\end{equation}
and the $\delta h^{ab}$ variation leads to
\begin{equation}
\partial _{a}\,X^{\mu }\partial _{b}X_{\mu }=\frac{1}{2}h_{ab}(h^{cd}%
\partial _{c}X^{\mu }\partial _{d}X_{\mu })  \label{h variation}
\end{equation}
All of these results have been shown above to follow from the conformal
dynamics of the integral transport curves of a self-dual $(1,2)$ tensor $%
P_{(0)}^{\mu }(x,y)$ (see eqs.(\ref{String wave eq.}), (\ref{Q}) and (\ref
{World sheet metric}), respectively). The mapping $\Phi ^{\mu }(x,y)$ is
necessarily causal because the Lorentz structure on $\mathcal{M}$ is
inherited by $\mathcal{N}_{\pi }$, and $\Phi ^{\mu }(x,y)$ is an immersion
because $P^{\mu }$ was assumed to be $C^{\infty }$ and $P^{\mu }$ and $%
Q^{\mu }$ are independent throughout $\mathcal{N}_{\pi }.$
\end{description}

\smallskip

Loosely stated, the theorem tells us that any measurable $(1,2)$ tensor
field corresponds to a string solution. Notice the important role played by
measurability in the proof. It was the use of self-dual equivalence classes
of operators that allowed us to show that $P^{\mu }Q_{\mu }=0,$ from which
the immersion followed. The metric structure on $\mathcal{N}_{\pi }$ follows
from the immersion. Finally, recall that the importance of using self-dual
equivalence classes is that self-duality provides the positive definiteness
needed to insure convergence of vectors on the tangent tower. In the final
summary section, we review this series of results.

\section{Summary}

The conformal dynamics of the weight tower associated with any
scale-invariant geometry follows directly from a careful treatment of the
classical physical principles of scale invariance. The mathematical
structures that result from this treatment include a Hilbert space
associated with each point of spacetime, as well as two of the most
important operator algebras of modern quantum field theory. While there
remain many implications of the new formalism to explore, and there are
sometimes puzzling differences of representation, the direct emergence of
such fundamental elements of physical theory is an important finding. In
this final section, we provide a short summary of our major results.

We show how the freedom to choose units for spacetime fields, and our usual
assumptions about dimensionful fields, leads to a tangent tower structure
associated with any scale-invariant spacetime. Because this tangent tower is
an infinite dimensional vector space, we require some convergence criterion
in order to define it and in order to form finite, measurable scalars. The
criterion we use is convergence in a positive definite norm, which produces
a Hilbert space on the tower. In order to guarantee positive definiteness of
the norm, only conformally self-dual vectors are allowed.

We next consider mappings on the tangent tower. Since only conformal and
Lorentz scalars are measurable, it is necessary to restrict linear mappings
on the tangent tower to those of definite or semi-definite conformal weight.
Then, the maximally commuting complete Lie algebra of definite-weight
operators is the set of unit, shifted-diagonal matrices, while the maximally
non-commuting complete Lie algebra of definite-weight operators is the
Virasoro algebra.

We develop a continuous representation for weight maps and show that the
maps having either vanishing commutators or vanishing anticommutors with the
dilation generator, $D,$ together span the space of harmonic functions in $2$%
-dim.

A study of the equivalence relations respecting the tangent tower structure
showed the existence of two equivalence relations which can be used to
define distinct tower structures. The first, weak equivalence, is consistent
only with the fully commuting (up to central charges) Lie algebra of zero
conformal weight operators. The operators of this algebra will all be
simultaneously measurable. The second equivalence relation, called strong
equivalence, allows non-commuting operators of various conformal weights.
Since only zero weight objects are measurable, objects on a strongly
equivalent tangent tower cannot be simultaneously measured.

We continue with the study of strong equivalence, showing that projective
representations of both the mode and Virasoro algebras identified earlier
occur with central charges. The projective representations are permitted by
both the weak and strong equivalence relations.

Moving to a simple particle example to study conformal dynamics, we show
that the parallel transport of an $(r,1)$ tensor satisfies the
Schr\"{o}dinger equation. The operator character of the Schr\"{o}dinger
equation follows because the spacetime connection acts on the entire tangent
tower. The factor of $i$ in the Schr\"{o}dinger equation arises because the
natural geometric variables (with $0$-signature $8$-dim metric) differ by a
factor of $i$ from the usual physical position and momentum variables (with
two $+2$-signature $4$-dim metrics). The model involves the identification
of the Hamiltonian operator with the time component of the dilational gauge
vector times the dilation operator. This interpretation has been
consistently implemented elsewhere [2],[3],[5].

Paralleling the conformal dynamics of a particle, we find the integral
curves of a general timelike, self-dual $(1,2)$ tensor. To integrate the
operator part of the field we demand that the field parallel transports the
vectors on which it acts. The resulting mapping from the $2$-parameter space
$\mathcal{N}_{\pi }=(0,\pi )\times (0,\pi )$ into spacetime is shown to be
an immersion of the same form as the open string solution. This immersion
induces a Lorentzian metric on $\mathcal{N}_{\pi }$ and satisfies the
variational equations governing open string.

In addition, a series of appendices develops a range of properties of the
new conformal formalism, including alternative representations for tower
tensors, the quantum algebra of weight operators, the use of length scales,
and many results on general and self-dual operators.

Taken as a whole, our results show that scale invariance provides an
underlying, principled reason for the physical importance of Hilbert space,
the Virasoro algebra, the string mode expansion, canonical commutators and
Schr\"{o}dinger evolution of states, all of which is independent of the
insights of both string theory and quantum theory.

\bigskip

\bigskip

The author is grateful for fruitful discussions with many physicists,
including P.G.O. Freund, P. M. Ho, G. Horowitz, E.
Martinec, C. G. Torre, and Y. S. Wu.

\pagebreak

\paragraph{Appendix 1: Alternate bases}

\smallskip

The simple form of eq.(\ref{D in def wt basis}) for $D(x,y)$ leads to a set
of definite weight functions $f_{k}(x)$ with only pure modes $f_{k}(x)\sim
e^{ikx}$, and definite weight operators $M_{(k)}(x,y)$ which for $k\neq 0$
have only ``right moving modes'', $M_{(k)}(x,y)\sim M_{(k)}(x-y).$ Such
heterosis [1] appears to be a necessary concomitant of definite weight
fields. To emphasize this fact, we repeat in this Appendix the calculation
of Sec.(5), starting with a more general ansatz for the form of $D(x,y).$ As
expected, we find that while most choices of basis lead to both right- and
left-moving modes, there is always a fixed relationship between the modes so
that there remains the same number of degreees of freedom.

We begin by choosing the form of $D(x,y)$ to be an arbitrary superposition
of both left- and right-moving Fourier modes, plus a possible correction for
surface terms
\begin{equation}
D(x,y)=-i\frac{\partial }{\partial x}(a\delta (x-y)+b\delta (x+y))-ic(x,y)
\end{equation}
where
\begin{eqnarray}
c(x,y) &=&c_{1}\delta (x+\pi )\delta (y+\pi )+c_{2}\delta (x+\pi )\delta
(y-\pi ) \\
&&+c_{3}\delta (x-\pi )\delta (y+\pi )+c_{4}\delta (x-\pi )\delta (y-\pi )
\end{eqnarray}
We seek a family of definite weight bases $f_{k}(x)$ for a range of values
of $(a,b).$ Despite containing both left and right modes, we find that this
set of bases still displays heterosis through a coupling of the amplitudes
of the two different sets of modes.

Consider the action of $D(x,y)$ on a $(0,1)$ vector
\begin{equation}
f(x)=\sum\limits_{k=-\infty }^{\infty }c_{k}\exp ikx
\end{equation}
where, as before, $f(\pi )=f(-\pi ).$ Computing $Df$ without $c(x,y)$ we
find a surface term of the form
\begin{equation}
\delta (x-\pi )[iaf(\pi )+ibf(-\pi )]-\delta (x+\pi )[ibf(\pi )+iaf(-\pi )]
\end{equation}
which vanishes only if $a=\pm b$ and $f(\pi )=\mp f(-\pi ).$ The
condition is even more stingent if we also allow $D$ to act on $f$ from the
right, $fD$ . For both surface terms to vanish requires $f(\pm
\pi )=0.$ In combination with the eigenvalue relation, eq.(\ref{Eigenvalue
eq.}), $f(\pm \pi )=0$ is inconsistent.

Therefore, the singular correction term is again necessary in the definition
of $D$. To find $c(x,y),$ consider the right and left action of $D$ on the $%
r^{th}$ slot of a rank $(0,k)$ tensor,
\begin{equation}
(DT)_{r}\equiv \int\limits_{-\pi }^{\pi }dzD(x,z)T(x_{1,}x_{2},\ldots
x_{r-1},z,x_{r-2},,\ldots x_{k})
\end{equation}
and a similar expression for $(TD)_{r}.$ For $(DT)_{r}$ the integration by
parts gives a surface term of the form
\begin{eqnarray}
S.T. &=&-i\left[ (-a\delta (x-z)+b\delta (x+z))T(x_{1,}x_{2},\ldots z\ldots
x_{k})\right] _{-\pi }^{\pi }  \nonumber \\
&&-i\int\limits_{-\pi }^{\pi }c(x,z)T(x_{1,}x_{2},\ldots z\ldots x_{k})
\end{eqnarray}
which cancels provided
\begin{eqnarray}
c_{1}+c_{2} &=&a+b \\
c_{3}+c_{4} &=&-a-b
\end{eqnarray}
Similarly, the surface term in evaluating $(TD)_{r}$ vanishes provided
\begin{eqnarray}
c_{1}+c_{3} &=&b-a \\
c_{2}+c_{4} &=&a-b
\end{eqnarray}
Both conditions are satisfied if $D$ is of the form
\begin{eqnarray}
D(x,y) &=&-ia[\partial _{x}\delta (x-y)-\delta (x-\pi )\delta (y+\pi
)+i\delta (x+\pi )\delta (y-\pi )] \\
&&-ib[\partial _{x}\delta (x+y)-\delta (x-\pi )\delta (y+\pi ) \\
&&\quad \quad \quad \quad \quad \quad \quad \quad -\delta (x+\pi )\delta
(y-\pi )+2\delta (x-\pi )\delta (y-\pi )]  \label{D with both modes}
\end{eqnarray}
plus an arbitrary multiple of the purely surface term
\begin{eqnarray}
E &=&[\delta (x-\pi )\delta (y-\pi )-\delta (x-\pi )\delta (y+\pi ) \\
&&-\delta (x+\pi )\delta (y-\pi )+\delta (x+\pi )\delta (y+\pi )]
\end{eqnarray}
Since $ET=TE=0$ for any tensor $T$ (as long as $T(\pi )=T(-\pi )$) we can
simply drop the $E$ term.

With eq.(\ref{D with both modes}) for $D,$ the left and right action of $%
D(x,y)$ on the $r^{th}$ slot of any tensor is simply given by the integrated
part
\begin{eqnarray}
(DT)_{r} &=&-ia\frac{\partial T}{\partial z}\mid _{z=x_{r}}+ib\frac{\partial
T}{\partial z}\mid _{z=-x_{r}}  \label{DT(a,b)} \\
(TD)_{r} &=&ia\frac{\partial T}{\partial z}\mid _{z=x_{r}}+ib\frac{\partial T%
}{\partial z}\mid _{z=-x_{r}}  \label{TD(a,b)}
\end{eqnarray}
Using eqs.(\ref{DT(a,b)}) and (\ref{TD(a,b)}) we now find a complete set of
eigenfuntions and definite weight operators.

Substituting $D(x,y)$ into
\begin{equation}
Df^{(k)}=kf^{(k)}
\end{equation}
we find the differential equation
\begin{equation}
Df^{(k)}=-ia\frac{\partial f^{(k)}(z)}{\partial z}\mid _{z=x_{r}}+ib\frac{%
\partial f^{(k)}(z)}{\partial z}\mid _{z=-x_{r}}=kf^{(k)}(x)
\label{D on f = kf}
\end{equation}
which is solved by the set of functions
\begin{equation}
f^{(k)}(x)=A_{k}(e^{ikx}\cosh \beta /2-e^{-ikx}\sinh \beta /2\ )
\end{equation}
provided $a$ and $b$ are related by
\begin{eqnarray}
a &=&\cosh \beta  \\
b &=&\sinh \beta
\end{eqnarray}
Fixing the overall normalization of $D(x,y)$ provides us with a $1$%
-parameter family of representations for the dilation operator$.$ Of course,
the $\beta =0$ case reduces $D$ to the form of eq.(\ref{D in def wt basis})
considered previously. In the special case where $a=b$ the zero mode
solution may be any sine series
\begin{equation}
f^{(0)}(x)=\sum_{n=1}^{\infty }A_{n}\sin nx
\end{equation}
while for $a=b,$ $k\neq 0$ the only solutions are symmetric functions with
discontinuous derivatives at the origin.

For $(r,2)$ tensors we can compute the commutator,
\begin{equation}
\lbrack D,M](x,y)=(DM)_{1}-(MD)_{2}
\end{equation}
which allows us to define operators $M_{(k)}$ of definite weight by
demanding
\begin{equation}
\lbrack D,M_{(k)}]=kM_{(k)}
\end{equation}
Substituting from eqs.(\ref{DT(a,b)}) and (\ref{TD(a,b)}),
\begin{equation}
a\partial _{x}M(x,y)-b\partial _{z}M(z,y)\mid _{z=-x}+a\partial
_{y}M(x,y)+b\partial _{z}M(x,z)\mid _{z=-y}=ikM(x,y)
\end{equation}
Separation of variables, $M(x,y)=f(x)g(y)$ leads to a pair of equations of
the form of eq.(\ref{D on f = kf}), giving the immediate solution:
\begin{eqnarray}
M_{(k)}(x,y) &=&\sum_{m=-\infty }^{\infty }\alpha
_{m}^{k}\,f^{(k-m)}(x)f^{(m)}(y) \\
&=&e^{ikx}h_{k}(x-y) \\
&&+\sinh \beta /2[e^{ikx}(h_{k}(x-y)\sinh \beta /2-h_{k}(x+y)\cosh \beta
/2)\ +c.c.]
\end{eqnarray}
where
\begin{equation}
h_{k}(z)\equiv \sum_{m=-\infty }^{\infty }\alpha _{m}^{k}e^{-imz}
\end{equation}
Thus, while $M_{(k)}(x,y)$ in general has both right and left modes, the
coefficients are not independent. Instead, they depend on the single
function $h_{k}(z)$ with $z$ replaced by either $x+y$ or $x-y.$

For $\beta =0,$ so that $D(x,y)$ contains only $\delta (x-y),$ the
commutator reduces to the previous expression,
\begin{equation}
\left( \frac{\partial }{\partial x}+\frac{\partial }{\partial y}\right)
M(x,y)=ikM(x,y)
\end{equation}
with solution
\begin{equation}
M_{(k)}(x,y)=\sum_{m=-\infty }^{\infty }\alpha _{m}^{k}e^{ikx}e^{-im(x-y)}
\end{equation}
as expected.

\smallskip \newpage

\paragraph{Appendix 2: A t'Hooft-Virasoro quantum algebra of weighted
operators}

\smallskip

The operators $M_{(k)},N_{(k)},J_{(k)}$ and $K_{(k)}$ defined in Secs.(2)
and (3) together form a quantum algebra. Writing the operators in the form
\begin{eqnarray}
M_{(k)}^{1,0} &=&M_{(k)}=e^{ikx}\sum \alpha _{m}^{k}e^{im(x-y)} \\
M_{(k)}^{1,1} &=&N_{(k)}=e^{ikx}\sum \alpha _{m}^{k}e^{im(x+y)} \\
M_{(k)}^{0,0} &=&J_{(k)}=e^{ikx}\sum e^{im(x-y)} \\
M_{(k)}^{0,1} &=&K_{(k)}=e^{ikx}\sum e^{im(x+y)}
\end{eqnarray}
where $\alpha _{m}^{k}=m.$ The graded t'Hooft commutators\smallskip are
defined by
\begin{equation}
\lbrack M_{(m)}^{\alpha ,\beta },M_{(n)}^{\alpha ^{\prime },\beta ^{\prime
}}]\equiv M_{(m)}^{\alpha ,\beta }M_{(n)}^{\alpha ^{\prime },\beta ^{\prime
}}-(-)^{\alpha \beta ^{\prime }+\beta \alpha ^{\prime }}e^{2i(\beta ^{\prime
}m-\beta n)x}M_{(n)}^{\alpha ^{\prime },\beta ^{\prime }}M_{(m)}^{\alpha
,\beta }
\end{equation}
where $m\geq n$ and $\alpha ,\beta ,\alpha ^{\prime },\beta ^{\prime }\in
\{0,1\}.$ The algebra is then given by
\begin{eqnarray}
\lbrack M_{(m)},M_{(n)}] &=&(n-m)M_{(m+n)} \\
\lbrack M_{(m)},J_{(n)}] &=&nJ_{(m+n)} \\
\lbrack M_{(m)},N_{(n)}]_{tH}^{+} &=&(m+n)N_{(m+n)} \\
\lbrack M_{(m)},K_{(n)}]_{tH}^{+} &=&nK_{(m+n)} \\
\lbrack J_{(m)},N_{(n)}]_{tH} &=&mK_{(m+n)} \\
\lbrack J_{(m)},J_{(n)}] &=&0 \\
\lbrack J_{(m)},K_{(n)}]_{tH} &=&0 \\
\lbrack N_{(m)},N_{(n)}]_{tH} &=&(n-m)M_{(m-n)} \\
\lbrack N_{(m)},K_{(n)}]_{tH}^{+} &=&-nJ_{(m-n)} \\
\lbrack K_{(m)},K_{(n)}]_{tH} &=&0
\end{eqnarray}
where a subscript $tH$ indicates the presence of a non-zero
phase, $e^{2i(\beta ^{\prime }m-\beta n)x},$ and a superscript $+$ denotes
an anticommutator due to the factor $(-)^{\alpha \beta ^{\prime }+\beta
\alpha ^{\prime }}$. Notice that $(M_{(k)},J_{(k)})$ form a Lie algebra, as
do $(J_{(0)},K_{(0)},M_{(0)},N_{(0)}).$

\pagebreak

\paragraph{Appendix 3: Length scales and phases}

A $(0,1)$ vector in the continuous representation is given by a Fourier
series
\begin{equation}
v(x)=\sum_{m=-\infty }^{\infty }v^{k}e^{ikx}
\end{equation}
Whereas the usual definite-weight quantities of physical theory have only a
single term, these vectors are superposition of all possible weights. The
question then arises, how do we make the connection with our usual units? In
this Appendix we address this question.

Measurements are made relative to some standard of length such as a meter
stick or a clock. Given a set of meter sticks (of length $l$) distributed
conveniently throughout spacetime, we construct the weight vector $%
l_{(1)}=l_{(1)}(x^{\mu }).$ In the continuous representation $l_{(1)}$
becomes $l_{1}(x^{\mu };x)=l_{(1)}e^{ix}.$ This field may be used to produce
a measurable length from a general tower vector $v(x)$ by taking the inner
product. This yields the dimensionless ratio
\begin{equation}
\left\langle v,l_{(1)}\right\rangle =\sum v^{n}l_{n}=\frac{v^{1}}{l}
\end{equation}
which is simply the magnitude of $v^{1}$ in units of $l.$

A related $(0,1)$ field,
\begin{equation}
\mathcal{L}(x)=\sum (l)^{n}e^{inx}
\end{equation}
in which the components $l^{n}$ are simply powers of $l,$ can be used to
interchange all of the phases for the usual units in a general vector $v(x).$
The inner product is the dimensionless expression
\begin{equation}
\left\langle v,l\right\rangle =\sum v^{n}l_{n}=\sum \frac{v^{n}}{(l)^{n}}
\end{equation}
in which the $(l)^{n}$-dimensioned component $v^{n}$ of $v$ is given in
units of $(l)^{n}.$

It is interesting to speculate about the consequences of conformal evolution
of $l_{1}(x).$ Presumably, $l(x^{\mu };x)$ will evolve by parallel transport
like any other weight vector. But we have no absolute knowledge of the
function $l_{1}(x^{\mu };x),$ rather, $l_{1}(x^{\mu };x)$ is our \textit{%
definition} of unit weight. Thus, even if $l_{1}$ evolved into a more
general superposition
\begin{equation}
l(x)=\sum l^{n}e^{inx}
\end{equation}
any measurement would treat it as a single mode
\begin{equation}
L(x)e^{iy(x)}\equiv \sum l^{n}e^{inx}
\end{equation}
effectively defining a new continuous parameter $y=y(x).$ Such repeated
redefinitions of the basis may account for ``collapse of the wave function.''

\pagebreak

\paragraph{Appendix 4: Vectors and their associated operators; properties of
self-dual operators}

There are certain $1-1$ maps which hold between vectors, definite weight
operators, and self-dual operators. These maps, and their generalizations to
$(r,1)$ tensors, play an important role in the relationship of conformal
dynamics to quantum dynamics. After studying these maps, we look in
particular detail at self-dual operators.

The first map, $\Psi ,$ relates any any $(0,1)$ vector, $v(x),$ to a
corresponding $(0,2)$ operator. We can use the identity operator
\begin{equation}
J_{0}=\sum e^{im(x-y)}=2\pi \delta (x-y)
\end{equation}
multiplicatively to produce the associated operator
\begin{equation}
V(x,y)=v(x)J_{0}
\end{equation}
Writing the expansion $v(x)=\sum \alpha ^{m}e^{imx}$ we see that producing $%
V(x,y)$ amounts to the replacement of the weight-$m$ vector $e^{imx}$ by the
weight-$m$ commuting operator $J_{m}$ since
\begin{equation}
V(x,y)=v(x)J_{0}=\sum_{m}(\alpha
^{m}e^{imx})\sum_{n}e^{in(x-y)}=\sum_{m}\alpha ^{m}J_{m}
\end{equation}
Starting with classical spacetime variables, it is easy to write down
corresponding tower vectors. $\Psi $ is then used to relate these vectors in
the self-dual Hilbert space to classical, commuting operators, which in turn
become non-commuting when the algebra acquires the central extension allowed
by projective representations.

A second pair of correspondences produces a $0$-weight operator and a
self-dual operator simply by varying the argument of $v(x).$ Unlike the
previous map, these correspondences do not preserve the conformal weight
associated with the coefficents of $\alpha ^{m}$, though they do still
amount to substitutions of an operator basis for the vector basis. For the $%
0 $-weight map the replacement
\begin{equation}
v(x)\rightarrow V_{(0)}(x,y)\equiv v(x-y)=\sum_{m}\alpha ^{m}e^{im(x-y)}
\end{equation}
gives a zero-weight operator, while a self-dual operator arises by replacing
$e^{imx}$ by the self-dual basis operator $\cos mx\ e^{-imy}$,
\begin{equation}
V_{SD}(x,y)=\frac{1}{2}(v(x-y)+v(-(x+y))=\sum_{m}\alpha ^{m}\cos mx\ e^{-imy}
\end{equation}

\smallskip Each of these $1-1$ relationships between vectors and operators
allows us to define an inner product on the operators via the vectors,
defining
\begin{equation}
\left\langle V,W\right\rangle =\left\langle \Psi v,\Psi w\right\rangle
\equiv \left\langle v,w\right\rangle
\end{equation}
for $\Psi $ and
\begin{equation}
\left\langle V_{(0)},W_{(0)}\right\rangle \equiv 2\left\langle
V_{SD},W_{SD}\right\rangle \equiv \left\langle v,w\right\rangle =\frac{1}{2%
\pi }\int\limits_{-\pi }^{\pi }\bar{v}^{*}(x)w(x)dx=\sum (c^{-n})^{*}d^{n}
\end{equation}
on the $0$-weight and self-dual operators by using the vector inner product.

Elementary calculations show how to write this invariant directly in terms
of integrals of the various operators. For $\Psi ,$ the inner product may be
expressed as the full parameter space integral
\begin{equation}
\left\langle V,W\right\rangle =\frac{1}{4\pi ^{2}}\int \int \bar{V}^{\dagger
}W\ dxdy
\end{equation}
while for the $0$-weight and self-dual correspondences, the result is a
trace
\begin{equation}
\left\langle V,W\right\rangle =tr(\bar{V}^{\dagger }W)=\left( \frac{1}{2\pi }
\right) ^{2}\int\limits_{-\pi }^{\pi }\bar{V}^{*}(x,y)W(x,y)dxdy
\end{equation}
In these expressions the adjoint dagger is the complex conjugate transpose, $%
V^{\dagger }(x,y)=V^{t*}(x,y)=V^{*}(y,x),$ while the bar denotes the
conformal dual. The conformal dual of an operator, $\bar{V},$ is defined so
that the dual vector $\overline{Vf}$ is the product of the duals, $\bar{V}%
\bar{f}.$ It follows that for both weight superpositions and self dual
operators the dual is found by replacing $\alpha ^{m}$ by $\alpha ^{-m},$%
\begin{eqnarray}
\bar{V}_{(0)}(x,y) &=&\sum \alpha ^{-m}\ e^{im(x-y)} \\
\bar{V}_{SD}(x,y) &=&\sum \alpha ^{-m}\cos mx\ e^{-imy}
\end{eqnarray}

The form $\bar{V}^{\dagger }W$ may be inferred directly by considering the
inner product of the vectors $Vf$ and $Wg$ for arbitrary $f$ and $g.$ Thus,
\begin{eqnarray}
\left\langle Vf,Wg\right\rangle &=&\left( \frac{1}{2\pi }\right)
^{3}\int\limits_{-\pi }^{\pi }\int\limits_{-\pi }^{\pi }\int\limits_{-\pi
}^{\pi }\overline{\left( V(x,y)f(y)\right) }^{*}\left( W(x,z)g(z)\right)
dxdydz \\
&=&\left( \frac{1}{2\pi }\right) ^{3}\int\limits_{-\pi }^{\pi
}\int\limits_{-\pi }^{\pi }\int\limits_{-\pi }^{\pi }\bar{V}^{*}(x,y)\bar{f}%
^{*}(y)\left( W(x,z)g(z)\right) dxdydz \\
&=&\left\langle f,\bar{V}^{\dagger }Wg\right\rangle
\end{eqnarray}

We can extend the inner product on self-dual operators to other operators as
well. For anti-self-dual operators, the conformal dual is also given by $%
\bar{V}(\alpha ^{m})=V(\alpha ^{-m}),$ but now the norm $\left\langle
Vf,Vf\right\rangle $ is negative for every $f$. To maintain positivity of
the norm for anti-self-dual operators, we therefore assign a multiplicative
grading, $\eta =0$ for self-dual and $\eta =1$ for anti-self-dual operators,
and write
\begin{equation}
\left\langle V,W\right\rangle =(-)^{\eta _{V}\eta _{W}}tr(\bar{V}^{\dagger
}W)
\end{equation}
where
\begin{equation}
\bar{V}=(-)^{\eta _{V}}V
\end{equation}

Finally, we can consider the inner product of a self-dual with an
anti-self-dual operator. Here the grading proves irrelevant, since the
product always vanishes
\begin{eqnarray}
\left\langle V_{SD},W_{ASD}\right\rangle  &=&(-)^{0\cdot 1}tr(\bar{V}
_{SD}^{\dagger }W_{ASD}) \\
&=&\left( \frac{1}{2\pi }\right) ^{2}\int\limits_{-\pi }^{\pi }\sum (\alpha
^{-m})^{*}\cos mx\ e^{imy}\sum (\beta ^{n})\sin nx\ e^{-iny}dxdy \\
&=&0
\end{eqnarray}
Thus, while the Hilbert norm is defined only for self-dual $(0,1)$ vectors,
the norm of $(r,2)$ tensors may be extended to include products of either
self-dual or anti-self-dual objects. This extension is useful for $r>0$
because it allows us to consistently define the Lorentz inner product
without requiring an immersion of the parameter space into spacetime. For
example, to determine whether the $(1,2)$ tensor
\begin{equation}
V^{\mu }(x,y)
\end{equation}
represents a timelike, null or spacelike vector we need to make a scalar of
the product
\begin{equation}
V^{\mu }(x,y)V_{\mu }(u,v)
\end{equation}
and it is important to do this in a way that doesn't bias the signs. The
grading above accomplishes this because
\begin{eqnarray}
\left\langle V_{SD}f,V_{SD}f\right\rangle  &=&\left\langle f,\bar{V}%
_{SD}^{\dagger }V_{SD}f\right\rangle \geq 0 \\
\left\langle V_{ASD}f,V_{ASD}f\right\rangle  &=&\left\langle f,\bar{V}
_{ASD}^{\dagger }V_{ASD}f\right\rangle \leq 0
\end{eqnarray}
for all vectors $f$. It is important to note that the inner product of
operator-valued vectors which are not orthogonal in spacetime may nontheless
vanish due to orthogonality of the operators. Thus, some familiar mnemonics
such as the nonvanishing of the inner product of any two timelike vectors
will no longer hold when such vectors are operator-valued.

A general operator may be decomposed into orthogonal parts
\begin{equation}
R=\frac{1}{2}(\bar{R}+R)+\frac{1}{2}(R-\bar{R})=R_{SD}+R_{ASD}
\end{equation}
The inner product of two such operators
\begin{equation}
\left\langle R,S\right\rangle =\left\langle R_{SD},S_{SD}\right\rangle
-\left\langle R_{ASD},S_{ASD}\right\rangle
\end{equation}
is then positive definite, $\left\langle R,R\right\rangle \geq 0.$ There is
no inherent conflict between the positivity of this norm on operators and
the existence of maps to the indefinite norm vector space, described at the
start of this Appendix. There are other maps possible which preserve the
positive norm. For example, we can map a general operator of the form
\begin{equation}
R=R_{SD}+R_{ASD}=\sum (\alpha ^{m}\cos mx\ +\beta ^{m}\sin mx)\ e^{-imy}
\end{equation}
to the self-dual vector
\begin{equation}
r=\sum \left( \alpha ^{m}\cos 2mx+\beta ^{m}\cos (2m+1)x\right)
\end{equation}
so that both $R$ and $r$ have positive norm. This map is clearly $1-1.$

Next, we consider the self-dual operator norm on equivalence classes of
self-dual operators. The equivalence classes exist because there are many
self-dual mappings that act in the same way on all self-dual vectors.
Unfortunately, varying over the set of operators equivalent in their effect
on self-dual vectors also varies the operator norm. The problem may be
resolved by maximizing the norm over the equivalence class.

Consider the action of two self-dual operators
\begin{eqnarray}
U &=&\sum \alpha ^{m}\cos mx\ e^{-imy} \\
V &=&\sum \beta ^{m}\cos mx\ e^{-imy}
\end{eqnarray}
on a self-dual vector
\begin{equation}
f=\sum c^{k}e^{ikx}
\end{equation}
where $c^{k}=c_{k}.$ The result is
\begin{equation}
Uf=\sum \alpha ^{m}c^{m}\cos mx=\alpha ^{0}c^{0}+\frac{1}{2}\sum_{m\neq
0}(\alpha ^{m}+\alpha _{m})c^{m}e^{imx}
\end{equation}
for $U$, and a similar result for $Vf.$ Equating $Uf$ and $Vf$ gives
\begin{equation}
\alpha ^{0}c^{0}+\frac{1}{2}\sum_{m\neq 0}(\alpha ^{m}+\alpha
_{m})c^{m}e^{imx}=\beta ^{0}c^{0}+\frac{1}{2}\sum_{m\neq 0}(\beta ^{m}+\beta
_{m})c^{m}e^{imx}
\end{equation}
or, since equality must hold for all convergent sequences $c^{m},$%
\begin{equation}
\alpha ^{m}+\alpha _{m}=\beta ^{m}+\beta _{m}\qquad \forall m
\end{equation}
We can parameterize the entire class by introducing a parameter $\lambda
_{m}\in [0,1]$. Then, starting with the member of the class having $\alpha
_{m}=0$ for all $m\neq 0,$ write
\begin{eqnarray}
\beta ^{m} &=&\lambda _{m}\alpha ^{m} \\
\beta _{m} &=&(1-\lambda _{m})\alpha ^{m}
\end{eqnarray}
while for $m=0$ we always have $\alpha ^{0}=\beta ^{0}.$ The norm of a
generic member of the equivalence class is then
\begin{equation}
\left\langle V,V\right\rangle =(\alpha ^{0})^{*}\alpha ^{0}+\frac{1}{2}%
\sum_{m\neq 0}\lambda _{m}(1-\lambda _{m})(\alpha ^{m})^{*}\alpha ^{m}
\end{equation}
We take the maximum of this norm as defining of the norm for the class. The
maximum is found by extremizing with respect to $\lambda _{m}$ for each $m$%
\begin{equation}
0=\frac{\partial \left\langle V,V\right\rangle }{\partial \lambda _{m}}%
=(1-2\lambda _{m})(\alpha ^{m})^{*}\alpha ^{m}
\end{equation}
so that
\begin{equation}
\lambda _{m}=\frac{1}{2}
\end{equation}
The norm of the class is therefore given by the norm of
\begin{equation}
V_{\max }=\beta ^{0}+\sum_{m\neq 0}\beta ^{m}\cos mx\ e^{-imy}
\end{equation}
where $\beta _{m}=\beta ^{m},$ given by
\begin{equation}
\left\langle V_{\max },V_{\max }\right\rangle =(\beta ^{0})^{*}\beta ^{0}+%
\frac{1}{2}\sum_{m\neq 0}(\beta ^{m})^{*}\beta ^{m}
\end{equation}
This norm has the convenient property that each term is positive definite.
In dealing with self-dual operator norms, we will always choose the
representative $V=V_{\max }$.

As a simple example of the operator inner product, we compute the inner
products of the $(1,2)$ tensors $P^{\mu }$ and $Q^{\mu }$ of Sec.(8). Noting
that $Q^{\mu }$ is anti-self-dual while $P^{\mu }$ is self-dual, we compute
the possible inner products in accordance with the rules developed above:
\begin{eqnarray}
\left\langle Q^{\mu },Q^{\nu }\right\rangle  &=&-tr((\bar{Q}_{\max }^{\mu
})^{\dagger }Q_{\max \ \mu }) \\
&=&-\frac{1}{4\pi ^{2}}\int \int dxdy\left[ \bar{Q}_{\max }^{\mu
}(x,y)\right] ^{*}Q_{\max \ \mu }(x,y) \\
&=&-\frac{1}{4\pi ^{2}}\int \int dxdy\overline{\left( \sum_{m\neq 0}\alpha
_{m}^{\mu }\sin (mx)\ e^{-imy}\right) }^{*}\left( \sum_{n\neq 0}\alpha _{\mu
n}\sin (nx)\ e^{-iny}\right)  \\
&=&-\frac{1}{2\pi }\int dx\sum_{m\neq 0}(\alpha ^{\mu m})^{*}\alpha _{\mu
}^{-m}\sin ^{2}(mx) \\
&=&-\frac{1}{2}\sum_{m\neq 0}(\alpha ^{\mu m})^{*}\ \alpha _{\mu m} \\
&=&-\frac{1}{2}\sum_{m\neq 0}(\alpha ^{\mu m})^{*}\ \alpha _{\mu }^{m}
\end{eqnarray}
where the final step follows because the norm is given by that of $Q^{\mu
}=Q_{\max }^{\mu }$. For the norm of the self-dual operator $P^{\mu }$ we
find
\begin{eqnarray}
\left\langle P^{\mu },P^{\nu }\right\rangle  &=&tr((\bar{P}_{\max }^{\mu
})^{\dagger }P_{\max \ \mu }) \\
&=&(\alpha ^{\mu 0})^{*}\ \alpha _{\mu 0}+\frac{1}{2}\sum_{m\neq 0}(\alpha
^{\mu m})^{*}\ \alpha _{\mu }^{m}<0
\end{eqnarray}
where the inequality follows because $P^{\mu }$ is timelike. Finally,
\begin{equation}
\left\langle P^{\mu },Q^{\nu }\right\rangle =0
\end{equation}
since $P^{\mu }$ and $Q^{\mu }$ have opposite duality.

This provides an example of the case alluded to above, where both of the $%
(1,2)$ vectors $u^{\mu }=\alpha _{0}^{\mu }$ and $v^{\mu }=P^{\mu }-\alpha
_{0}^{\mu }$ are timelike, $u^{\mu }u_{\mu }=\alpha _{0}^{\mu }\alpha _{\mu
0}<0$ and $v^{\mu }v_{\mu }=\sum_{m\neq 0}(\alpha _{m}^{\mu })^{*}\alpha
_{\mu m}<0,$ even though they are orthogonal, $u_{\mu }v^{\mu }=0.$ This
happens because $P^{\mu }=u^{\mu }+v^{\mu }$ is an orthogonal operator
decomposition.

We conclude with two particular types of self-dual maps: identity maps and
constant maps. Identity maps are those which map every self-dual vector $f$
to a multiple of $f,$%
\begin{equation}
J_{SD}f=\lambda f
\end{equation}
They are necessarily of the form
\begin{equation}
J_{SD}(x,y)=\frac{1}{2}\lambda (J_{0}+K_{0})
\end{equation}
There are also rank $(r,2)$ tensor valued identity maps $J_{SD},$ which take
the form
\begin{equation}
J_{SD}^{\mu \cdots \nu }(x,y)=\frac{1}{2}\lambda ^{\mu \cdots \nu
}(J_{0}+K_{0})
\end{equation}
Constant maps are those that map every vector $f$ to the same constant
value,
\begin{equation}
C_{SD}f=c
\end{equation}
These maps consist of the zero-mode only,
\begin{equation}
C_{SD}=\sum \alpha _{m}\cos mx\ e^{imy}=\alpha _{0}
\end{equation}
and also have genereralizations to $(r,2)$ tensors given by
\begin{equation}
C_{SD}^{\mu \cdots \nu }=\alpha _{0}^{\mu \cdots \nu }
\end{equation}
Notice that the norm of an equivalence class of self-dual maps $V_{SD}$ is
\begin{equation}
\left\langle V_{SD},V_{SD}\right\rangle =(\alpha ^{0})^{*}\alpha ^{0}
\end{equation}
if and only if $V$ is a constant map. For proof, we have that the norm of $%
V_{SD}$ is equal to the norm of $V_{\max },$ which is $(\alpha
^{0})^{*}\alpha ^{0}+\frac{1}{2}\sum_{m\neq 0}(\alpha ^{m})^{*}\alpha ^{m}.$
Since the norm is also $(\alpha ^{0})^{*}\alpha ^{0}$ and $(\alpha
^{m})^{*}\alpha ^{m}$ is positive definite, we must have $\alpha ^{m}=0$ for
all $m\neq 0$ in $V_{\max },$ so $V_{\max }=V_{SD}=\alpha ^{0}.$

\pagebreak

\paragraph{Appendix 5: Vector-valued operator algebras}

The maximally and minimally commuting complete Lie algebras of definite
weight operators were shown in Secs.(3) and (4) to be the mode and Virasoro
algebras, respectively. It is a simple matter to extend these algebras to $%
(1,2)$ tensor algebras of operators. Suppose we have definite weight
operators $G_{(m)}$ satisfying
\begin{equation}
\lbrack G_{(m)},G_{(n)}]=a_{mn}G_{(m+n)}+c_{mn}\mathbf{1}
\end{equation}
Now consider the possible corresponding algebra for definite weight $(1,2)$
tensors $H_{(m)}^{a}.$ Clearly
\begin{equation}
\lbrack H_{(m)}^{a},H_{(n)}^{b}]=a_{mn}T_{\quad c}^{ab}\
H_{(m+n)}^{c}+c_{mn}S^{ab}\mathbf{1}
\end{equation}
where $T_{\quad c}^{ab}$ and $S^{ab}$ are Lorentz tensors, symmetric on $ab$%
. These generalizations are only possible if the Lorentz part of the tangent
space under consideration has appropriate tensor fields $T_{\quad c}^{ab}$
and $S^{ab}.$ Unless the spacetime has nonvanishing torsion, there is no
rank-$3$ tensor available, but there is always the Lorentz metric for $%
S^{ab}.$ Therefore, the mode algebra generalizes immediately to
\begin{equation}
\lbrack J_{(m)}^{a},J_{(n)}^{b}]=m\delta _{m+n}^{0}\eta ^{ab}\mathbf{1}
\end{equation}
while there is no vector-valued generalization of the Virasoro algebra.

The existence of this algebra permits us to add a map from $(1,1)$ tensors
to $(1,2)$ tensors, parallel to the maps of $(0,1)$ tensors to $(0,2)$
tensors presented in Appendix 4. Starting with
\begin{equation}
v^{a}(x)=\sum v^{am}e^{imx}
\end{equation}
we can map to the superpostion of definite weight operators
\begin{equation}
V^{a}(x,y)=v^{a}(x)J_{0}(x,y)=\sum_{m,n}v^{am}e^{imx}e^{in(x-y)}=%
\sum_{m}v^{am}J_{m}(x,y)\equiv \sum_{m}V_{m}^{a}
\end{equation}
In the absence of central charges, the $V^{a}$ commute
\begin{equation}
\lbrack
V^{a},V^{b}]=\sum_{m,n}v^{am}v^{bn}(e^{imx}e^{iny}-e^{imy}e^{inx})%
\sum_{k}e^{ik(x-y)}=0
\end{equation}
where the sums vanish because $\sum_{k}e^{ik(x-y)}=2\pi \delta (x-y).$

Naturally the definite weight operators $V_{m}^{a}$ also commute, up to the
central charges allowed by projective representations. To treat the central
charges, we need to establish a basis to relate the coefficients $V_{m}^{a}$
to the ones in the definition of the $J_{m}^{a}$ algebra. Let a basis
algebra $J_{m}^{a}$ be defined by choosing a vielbein, $e_{m\mu }^{\quad a}$
for each $m.$ Then, for each $m$, we have commuting operators
\begin{equation}
J_{\mu (m)}^{a}=\sum_{n}e_{m\mu }^{\quad a}e^{imx}e^{in(x-y)}
\end{equation}
With the most general allowed central charges, the $J_{\mu (m)}^{a}$ satisfy
\begin{equation}
\lbrack J_{\mu (m)}^{a},J_{\nu (n)}^{b}]=m\delta _{m+n}^{0}(\lambda _{1}\eta
^{ab}g_{\mu \nu }+\lambda _{2}\varepsilon _{\quad cd}^{ab}\ e_{m\mu }^{\quad
c}\ e_{n\nu }^{\quad c})\mathbf{1}
\end{equation}
where for particular physical fields parity considerations may constrain $%
\lambda _{1}$ or $\lambda _{2}$ to vanish. In any case, we can expand
general vectors $V_{(m)}^{a}$ in terms of the orthonormal basis
\begin{equation}
V_{(m)}^{a}=V_{(m)}^{\mu }e_{m\mu }^{\quad a}
\end{equation}
Then the central charges,
\begin{eqnarray}
\lbrack V_{(m)}^{a},V_{(n)}^{b}] &=&V_{(m)}^{\mu }V_{(n)}^{\nu }[J_{\mu
(m)}^{a},J_{\nu (n)}^{b}] \\
&=&m(\lambda _{1}V^{\mu m}V_{\mu m}\eta ^{ab}+\lambda _{2}\varepsilon
_{\quad cd}^{ab}V_{(m)}^{c}V_{(-m)}^{d})\delta _{m+n}^{0}\mathbf{1}
\end{eqnarray}
are linear combinations of the combined conformally invariant and Lorentz
invariant magnitude $V^{\mu m}V_{\mu m}$ and of the invariant antisymetric
product $\varepsilon _{\quad cd}^{ab}V_{(m)}^{c}V_{(-m)}^{d}.$ The second
term vanishes for self-dual fields.

\pagebreak

\paragraph{References}

\smallskip

\begin{enumerate}
\item  Green, M.B., J. H. Schwarz and E. Witten, \textit{Superstring theory}
, Cambridge University Press (1987).

\item  Wheeler, J. T., $SU(3)\times SU(2)\times U(1)$\textit{: The residual
symmetry of conformal gravity}, Contemporary Mathematics, Vol. \textbf{132},
(1992) pp 635-644.

\item  Wheeler, J. T., \textit{Quanta without quantization}, Honorable
Mention Gravity Essay 1997, submitted for publication, hep-th/9705235.

\item  Wheeler, J. T., \textit{New conformal gauging and the electromagnetic
theory of Weyl}, submitted for publication.

\item  Weinberg, S., \textit{The quantum theory of fields,} Cambridge
University Press (1995).

\item  Wheeler, J.T., \textit{Quantum measurement and geometry}, Phys. Rev.
\textbf{D}41, 2, (1990) 431.
\end{enumerate}

\end{document}